\documentclass[preprint]{aastex}
\begin{document}
\def\NL{$2{\times}10^{20}$cm$^{-2}$}
\def\NH{$N$(HI)}
\def\fNX{$f(N,X)$}
\def\fperpNX{$g(N_{\perp},X)$}
\def\nperp{$N_{\perp}$}
\def\pkts{P_{KS}}
\def\etal{et al.}
\def\Lya{Ly$\alpha$ }
\def\lya{Ly$\alpha$ }
\def\smpy{M$_{\odot}{\rm \ yr^{-1} }$}
\def\smpykpc{M$_{\odot}{\rm \ yr^{-1} \ kpc^{-2}}$}
\def\smpympc{M$_{\odot}{\rm \ yr^{-1} \ Mpc^{-3}}$}
\def\lcdm{$\Lambda$CDM }
\def\rvec{${\bf r}$}
\def\Rratio{${{\cal R}(v)}$}
\def\ndmp{17}
\def\nrun{8,500}
\def\kms{km~s$^{-1}$ }
\def\omgm{$\Omega_{{\rm M}}$}
\def\omgv{$\Omega_{\Lambda} $}
\def\omgas{$\Omega_{g}(z) $}
\def\dv{$\Delta v$}
\def\thetkern{$\theta_{\rm kern}$}
\def\thetpsf{$\theta_{\rm PSF}$}
\def\thetexp{$r_{\rm s}$}
\def\dthets{$\Delta \theta_{\rm exp}$}
\def\thetdla{$\theta_{\rm dla}$}
\def\nh{$N$}
\def\thresh{2{$\times$}10$^{20}$ cm$^{-2}$}
\def\dvm{$\delta v$}
\def\vc{$V_{rot}(r)$}
\def\vrot{$V_{rot}(r)$}
\def\micron{$\mu$m}
\def\gamd{${\Gamma_{d}(\bf r)}$}
\def\gamdnr{${\Gamma_{d}}$}
\def\peq{$P_{eq}$}
\def\pmin{$P_{min}$}
\def\pmax{$P_{max}$}
\def\taunu{$\tau_{\nu}$}
\def\pgeom{$(P_{min}P_{max})^{1/2}$}
\def\kap{${\kappa ({\bf r})}$}
\def\dNdX{$d{\cal N}/dX$}
\def\kapnr{${\kappa}$}
\def\cm2{${\rm cm}^{-2}$}
\def\cm3{${\rm cm}^{-3}$}
\def\N#1{{N({\rm #1})}}
\def\f#1{{f_{\rm #1}}}
\def\rAA{{\rm \, \AA}}
\def\sci#1{{\rm \; \times \; 10^{#1}}}
\def\ltk{\left [ \,}
\def\ltp{\left ( \,}
\def\ltb{\left \{ \,}
\def\rtk{\, \right  ] }
\def\rtp{\, \right  ) }
\def\rtb{\, \right \} }
\def\ohf{{1 \over 2}}
\def\nohf{{-1 \over 2}}
\def\rhf{{3 \over 2}}
\def\smm{\sum\limits}
\def\perd{\;\;\; .}
\def\cmma{\;\;\; ,}
\def\semic{\;\;\; ;}
\def\sgint{\sigma_{int}}
\def\frat{$f_{ratio}$}
\def\intl{\int\limits}
\def\rhodot{$\dot{\rho_{*}}$}
\def\rhodotN{$\dot{\rho_{*}}({\ge}N)$}
\def\rhodotpsi{$\dot{\rho_{*}}({\ge}{\dot{\psi_{*}}})$}
\def\rhodotz{$\dot{\rho_{*}}$$(z)$}
\def\Mdot{$\dot{M_{*}}$}
\def\rhosz{${\rho_{*}(z)}$}
\def\rhos{${\rho_{*}}$}
\def\Nperp{${N_{\perp}}$}
\def\msolar{M$_{\odot}$}
\def\und#1{{\rm \underline{#1}}}
\def\ps{$\dot{\psi_{*}}$}
\def\psthresh{($\dot{\psi_{*}})_{\rm thresh}$}
\def\ms{$\dot{M_{*}}$}
\def\psav{$<$$\dot{\psi_{*}}$$>$}
\def\psavz{${<{{\dot{\psi_{*}}}}(z)>}$}
\def\ciis{C II$^{*}$}
\def\lc{${\ell}_{c}$}
\def\lclos{${\ell}_{c}$}
\def\lcav{$<{\ell}_{c}>$}
\def\lcr{$l_{cr}({\rm {\bf r}})$}
\def\lcrnr{$l_{cr}$}
\def\jnu{$J_{\nu}$}
\def\knu{$k_{\nu}$}
\def\junit{ergs cm$^{-2}$ s$^{-1}$ Hz$^{-1}$ sr$^{-1}$}
\def\arcsec{\hbox{$^{\prime\prime}$}}

\title{SEARCHING FOR LOW SURFACE-BRIGHTNESS 
GALAXIES IN THE HUBBLE ULTRA DEEP FIELD: IMPLICATIONS FOR 
THE STAR FORMATION EFFICIENCY IN NEUTRAL GAS AT $z\sim 3$
\altaffilmark{1}}

\author{ Arthur M. Wolfe\\ 
Department of Physics, and Center for Astrophysics and Space Sciences; \\
University of California, San
Diego; \\
9500 Gilman Dr.\\
La Jolla; CA 92093-0424\\
{\bf awolfe@ucsd.edu}}

\author{and}

\author{ Hsiao-Wen Chen\\ 
Department of Astronomy \& Astrophysics,  \\
University of Chicago; \\
Chicago, IL 60637 \\
{\bf hchen@oddjob.uchicago.edu}}

\altaffiltext{1}{Based on observations obtained with the NASA/ESA 
{\em Hubble Space Telescope}. 
{\em HST} is operated
by the Association of Universities for Research in 
Astronomy (AURA), Inc., under NASA contract NASS5-26555.}


\begin{abstract} The Kennicutt--Schmidt law relates the face-on star formation rate
(SFR) per unit area with the face-on gaseous column density in nearby
galaxies.  Applying this relation to damped {\lya} absorption systems
(DLAs) of neutral hydrogen column density {\nh} $> 1.6\times
10^{21}$ cm$^{-2}$ leads to an estimate that three percent of the sky
should be covered with extended sources brighter than $\mu_{V}\approx
28.4$ mag arcsec$^{-2}$, if DLAs at redshift $z$=[2.5,3.5] undergo {\em
in situ} star formation.  We test this hypothesis by searching the
Hubble Ultra Deep Field (UDF) F606W image for low surface-brightness
features of angular sizes, ranging between {\thetdla}=0.25\arcsec\ and
4.0\arcsec.  At $z=2.5-3.5$, the observed F606W fluxes correspond to
roughly rest-frame 1500 \AA\ and the angular sizes correspond to
predicted disk diameters of $d_{\rm dla}=2-31$ kpc.  After convolving
the F606W image with smoothing kernels of angular diameters
{\thetkern}={\thetdla}, we find the number of detected objects to
decrease rapidly to zero at {\thetkern}$>$1 {\arcsec}.
Our search yields  upper limits on the comoving SFR densities that
are between factors of 30 and 100 lower than predictions, suggesting a
reduction by more than a factor of 10
in star formation efficiency at $z\sim 3$.  We
consider several mechanisms that could reduce star formation
efficiency at high redshift. We find that the cosmological increase with
redshift of the critical surface density for the Toomre instability may be
sufficient to suppress star formation to the levels implied by the
UDF observations. However, the uncertainties are such that Toomre
instabilities may still exist. In that case star formation at column
densities less than 10$^{22}$ cm$^{-2}$ may be suppressed
by
the low molecular content of the DLA gas.
The upper limits on {\em in situ} star formation
reduce the predicted metallicities at $z\sim 3$ to be
significantly lower than observed, and reduce the heat input in the gas to be
substantially lower than the inferred cooling rates.  In contrast, the
radiative output from compact Lyman Break Galaxies (LBGs) 
with $R$ $<$ 27 is
sufficient to balance the comoving cooling rate.  This leads us to
posit that a significant fraction of the DLA population are hosts to
more compact regions
of active star formation, which may be the sources of metal
enrichment for these DLAs. Such regions are likely to be LBGs.

\end{abstract}



\section{INTRODUCTION}

A principal goal of galaxy formation theory is to understand how stars
form from gas.  This condensation process is fundamental for
determining the star-formation history of a galaxy, the spatial
distributions of its stellar populations, and its chemical
evolution. Yet it is still not understood.  On small scales, the
physical processes are complex and difficult to calculate (e.g. Shu
{\etal} 1987; Krumholz \& McKee 2005), while on cosmological scales,
realistic simulations of star formation are in addition constrained by
limits on numerical resolution (e.g. Cen {\etal} 2003; Nagamine
{\etal} 2004a, 2004b).  On the other hand considerable progress has
been made in the observational sector. Recent multi-color imaging and
spectroscopic surveys have succeeded in tracing starburst galaxies out
to redshifts as large as six (Giavalisco {\etal} 2004; Bouwens {\etal}
2004). The majority of galaxies found in this way are the Lyman
Break Galaxies (i.e. LBGs; e.g. Steidel {\etal} 2003).  These are
compact (half-light {\em diameters} $\sim$ 4 kpc) star-forming
galaxies with mean star formation rates SFR $\sim$ 40 {\smpy}, after
extinction corrections are applied (Shapley {\etal} 2003).
The LBGs contribute a SFR per unit comoving volume {\rhodotz}
(Steidel {\etal} 1999; Giavalisco {\etal} 2004)
that in the redshift interval $z$ =[6,2] would consume 
a mass per unit comoving volume of cold neutral gas
equivalent to about 10 $\%$ of the 
mass content of visible stars in modern galaxies,
i.e., 0.1{$\Omega_{*}(z=0)$}. As a result, 
reservoirs of neutral gas
at $z \ \ge$ 2 
may be required to fuel these star-forming objects. 

The purpose of this paper is to search for star formation in
spatially extended regions that serve as such neutral-gas
reservoirs. We focus on the damped {\lya} systems (hereafter DLAs),
the population of quasar absorption systems with {\nh} {$\ge$}
{\thresh}, where {\nh} denotes observed H I column density (for a
review see Wolfe, Gawiser, \& Prochaska 2005 [hereafter WGP05]).  At
$z$ $\sim$ 3 this column-density threshold guarantees gas neutrality
in most cases, which distinguishes DLAs from the {\lya} forest and all
other classes of absorption systems in which {\nh} $<$ {\thresh} and
the gas is more than 50 $\%$  
ionized.
The neutrality of the gas takes on added significance when it is
realized that DLAs (1) dominate the neutral-gas content of the
Universe at $z$ =[0,5] and (2) at $z$ $\approx$ 3.5 contain sufficient
gas to account for 0.5{$\Omega_{*}(z=0)$}. 
Furthermore surveys for DLAs reveal a large area-covering
factor for neutral gas with {\nh} {$\ge$} {\thresh} (Wolfe {\etal}
1995; Prochaska, Herbert-Fort, \& Wolfe 2005 [PHW05]).  In the redshift interval $z$=[2.5,3.5], a redshift range in which the SFR history
of galaxies is well determined,
DLAs cover one third of the sky, which leads to a striking
conclusion: the sky should be ``lit up'' with star
formation. Specifically, a subset of DLAs with $N$ $>$
1.6{$\times$}10$^{21}$ cm$^{-2}$ should cover three percent of the sky
and have surface brightnesses brighter than 28.4 mag arcsec$^{-2}$
(see below).  This follows if one presumes the Kennicutt--Schmidt law
(Schmidt 1959; Kennicutt 1998a,b) holds at high redshifts. In that case
star formation occurs in the presence of cold atomic and/or molecular
gas with a SFR per unit area projected perpendicular to the disk given
by

\begin{equation}
({\dot{{\psi}_{*}}})_{\perp}=\left\{
\begin{array}{ll}
0 \ ; N_{\perp} < N_{\perp}^{crit} \\
K{\times}[N_{\perp}/N_{c}]^{\beta}  \ ; N_{\perp} {\ge} N_{\perp}^{crit},
\end{array}
\right.
\label{eq:Kenlaw}
\end{equation}


\noindent where $N_{\perp}$ is the H I column density perpendicular to
the disk.  
In nearby galaxies, $K$=$K_{Kenn}$
=(2.5$\pm$0.5){$\times$}10$^{-4}$ {\smpykpc}, ${\beta}$=1.4$\pm$0.15,
and the scale factor $N_{c}$=1.25{$\times$}10$^{20}$ cm$^{-2}$
(Kennicutt 1998a,b). The threshold column density 
$N_{\perp}^{crit}$  is observed to range between 5{$\times$}10$^{20}$ cm$^{-2}$
and 2{$\times$}10$^{21}$ cm$^{-2}$ (Kennicutt 1998b) and is
usually associated with the threshold condition for the Toomre
instability. We ignore the molecular component of the gas owing 
to the very low H$_{2}$ area covering factor of DLAs (e.g. Ledoux {\etal} 2003).

In this paper, we test the hypothesis that star formation
proceeds throughout the absorbing gas at the projected rates given by
Eq.~\ref{eq:Kenlaw}. That is, we consider whether 
star formation at high $z$ occurs in sites other than 
compact LBGs. 
In $\S$ 2 we apply Eq.~{\ref{eq:Kenlaw}} to
high-$z$ DLAs and compute their expected surface brightnesses.
We compute for the first time the SFR per unit comoving
volume for DLAs modeled as randomly oriented disks.  
We then calculate the number of DLAs
expected to occupy the Hubble Ultra Deep Field (UDF; S. V. W. Beckwith {\etal}
2004, in preparation). 
In $\S$ 3 we
describe a search in the UDF for extended regions with low surface
brightnesses.  We use a matched kernel technique optimized for
detecting faint emission from extended objects and discuss the results of
our search in $\S$ 4. In $\S$ 5 we describe the implications of these results.
A summary and concluding remarks are given in $\S$ 6.

Throughout this paper we adopt a  cosmology with
(${\Omega_{\rm M}},{\Omega_{\Lambda}},h$)=(0.3,0.7.0.7) 
(Spergel {\etal} 2003). We adopt the 
AB magnitude system and refer to magnitudes and surface brightnesses 
deduced from the F606W image with the ACS camera on HST as follows:
$V$ $\equiv$ 
$AB({\rm F606W})$ and
{$\mu_{V}$} $\equiv$
$\mu_{\rm F606W}$.

\section{THEORETICAL FRAMEWORK}

Before describing the observational results and analysis
we discuss observational properties predicted for DLAs relevant
for detecting them in emission.

\subsection{Comoving SFR Densities Predicted by the Kennicutt--Schmidt Law}
In order to estimate the surface
brightnesses of DLAs with projected SFRs predicted by the
Kennicutt--Schmidt law, we assume they are disks inclined to the plane
of the sky by inclination angles $i$.
In that case the observed intensity at frequency $\nu_{0}$ is given by

\begin{equation}
I_{\nu_{0}}(i)={{({\Sigma_{\nu}})_{\perp}} \over {4{\pi}{(1+z)^{3}{\rm
cos}(i)}}}
\label{eq:Inui}
\cmma
\end{equation}

\noindent where ${({\Sigma_{\nu}} )_{\perp}}$ is the luminosity per
unit frequency interval per unit area projected 
perpendicular to the plane of the
disk and ${\nu}=(1+z){\nu_{0}}$: owing to the low dust-to-gas ratios
of most DLAs (Pettini 2004), we ignore the effects of extinction.  In the
case of FUV radiation (with $\lambda$ $\approx$ 1500 {\AA}),
$({\Sigma_{\nu}})_{\perp}$=$C$ ({\ps})$_{\perp}$ where
$C$=8{$\times$}10$^{27}$ ergs s$^{-1}$ Hz$^{-1}$({\msolar}
yr$^{-1}$)$^{-1}$/(3.08{$\times$}10$^{21}$ cm/kpc)$^{2}$=
8.4{$\times$}10$^{-16}$ ergs cm$^{-2}$
s$^{-1}$Hz$^{-1}$({\smpykpc})$^{-1}$,
where the calibration is insensitive to
wavelength in the FUV portion of the
spectrum (Madau, Pozzetti, \& Dickinson 1998).  Because
$N_{\perp}$=cos($i$)$N$, we find that, for a fixed
observed column density, $I_{\nu_{0}}(i)$ averaged over
0$^{o}$ $\le$ $i$ $\le$ 90$^{o}$
is given by

\begin{equation}
<I_{\nu_{0}}>={C{\dot{{\psi}_{*}}} \over {4{\pi}{(1+z)^{3}{\beta}}}} 
\cmma {\dot{{\psi}_{*}}{\equiv}K(N/N_{c})}^{\beta}
\label{eq:Inuav}
\perd
\end{equation}

\noindent Note that {\ps} is {\it not} the average SFR per unit area
projected along the line of sight,$<$({\ps})$_{\perp}$/cos($i$)$>$.
Rather it is an effective projected SFR relating the observed column
density, $N$, to the surface brightness of the DLA. As an example,
consider a DLA with {$N$} =1.6{$\times$}10$^{21}$ cm$^{-2}$, which is
about twice the mean value of {$N$} for the statistical sample of over
625 DLAs (PHW05). From
Eq. {\ref{eq:Inuav}} we find {\ps}=8.9{$\times$}$10^{-3}$
{\smpykpc} (about twice
the local rate in the Galaxy). 
At $z$ = 3 this corresponds to an $AB$ surface brightness
in the $V$ band, $\mu_{V}$= 28.4 mag arcsec$^{-2}$.  Here we assumed
the values for $K, {\beta}, {\rm and} \ {N_{c}}$ cited above (Kennicutt
1998a,b).  While objects this faint are beyond the sensitivity of the 
Hubble Deep Field (HDF)
and the GOODs survey, they are within the sensitivity of images
acquired by the UDF (Bouwens {\etal} 2004).

We next compute the comoving SFR density
{\rhodotz} predicted by the Kennicutt--Schmidt law in order to compare
it with the empirical values determined from the UDF (see $\S$ 3). The
calculation was originally carried out by Lanzetta {\etal} (2002: see
also Hopkins {\etal} 2005). Here for the first time we account for
inclination effects that must be present if, as we assume, DLAs are
disk-like structures or any type of gaseous configurations with
preferred planes of symmetry, such as those predicted
in high-resolution numerical simulations (e.g. Razoumov {\etal} 2005; Kravtsov 2003).
For disks with inclination angles
between $i$ and $i+di$ the comoving SFR density due to the intrinsic
differential area $dA_{\perp}$ of the disk is given by

\begin{equation}
d^{2}{\dot{\rho_{*}}}=n_{co}{dA_{\perp}}{\rm cos}(i){\times}[{({\dot{\psi_{*}}})_{\perp}}/{\rm cos}(i)]{\times}{\rm sin}(i)di
\label{eq:drhostardot}
\cmma
\end{equation}

\noindent where $n_{co}$ is the comoving density of disks.  Noting
that the intrinsic column-density distribution of the disk is
defined by the relation

\begin{equation}
g(N_{\perp},X)dN_{\perp} \equiv(c/H_{0})n_{co}(X)dA_{\perp} 
\label{eq:gdN}
\cmma
\end{equation}

\noindent and cos($i$)=$N_{\perp}/N$ and sin($i$)$di$=($N_{\perp}/N^{2})dN$
we
find that

\begin{equation}
{{\dot{\rho_{*}}}({\ge}N,X)}=(H_{0}/c){\int_{N}^{N_{max}}dNK(N/N_{c})^{\beta}}{{\int_{N_{min}}^{{\rm min}(N_{0},N)}dN_{\perp}g(N_{\perp},X)}}(N_{\perp}^{2}/N^{3})(N_{\perp}/N)^{{\beta}-1}
\label{eq:rhostardotN}
\cmma
\end{equation}

\noindent where $g$({\Nperp},$X$) and the observed column-density
distribution function {\fNX} are related by

\begin{equation}
f(N,X)={{\int_{N_{min}}^{{\rm min}(N_{0},N)}dN_{\perp}g(N_{\perp},X)}}(N_{\perp}^{2}/N^{3})
\label{eq:fNNperp}
\end{equation}

\noindent 
(Fall \& Pei 1993; Wolfe {\etal} 1995)
and $X(z)$ is the absorption distance (Bahcall \& Peebles 1969). 
In deriving these equations we
implicitly assumed {\Nperp} to be a monotonically decreasing function
of radius with a maximum value of $N_{0}$. Note, in the spherically
symmetric limit the $(N_{\perp}/N)^{{\beta}-1}$ term in
Eq.~{\ref{eq:rhostardotN}} would be replaced by 1 and consequently the
expression for {\rhodot}(${\ge}N,X$) would be the same as the 
Lanzetta {\etal} (2002) result (see also Hopkins {\etal} 2005).

Equations~{\ref{eq:rhostardotN} and {\ref{eq:fNNperp}} show that one
must obtain {\fperpNX} from {\fNX} to compute {\rhodot}(${\ge}N,X$). The double
power-law fit to the SDSS data resulted in {\fNX}
=$k_{3}$$(N/N_{d})^{\alpha}$ where $k_{3}$
=(1.48$\pm$0.07}){$\times$}10$^{-24}$ cm$^{2}$,
$\alpha$=$\alpha_{3}$=$-$2.00$\pm$0.06 for $N$ ${\le}$ $N_{d}$ and
$\alpha$=$\alpha_{4}$ =$-$6.00$^{+4.06}_{-3.93}$ at $N$ $>$ $N_{d}$,
where $N_{d}$=(3.16$^{+0.47}_{-0.28}$){$\times$}10$^{21}$ cm$^{-2}$
(PHW05).  It is natural to equate $N_{d}$ with $N_{0}$ in which case
Eq.~{\ref{eq:fNNperp}} shows that projection effects alone result in
$\alpha_{4}$= $-$3.0. This solution is clearly consistent with the
data, given the large uncertainties in $\alpha_{4}$
(note, {\rhodot}[$\ge N,X$] is independent of $\alpha_{4}$). As a result this
fit leads to {\fperpNX}= $k_{3}$$(N_{\perp}/N_{0})^{-2}$ at
$N_{\perp}$ $\le$ $N_{0}$ and {\fperpNX}=0 for $N_{\perp}$ $>$
$N_{0}$. One could envisage an alternative solution in which $N_{0}$
$\sim$ 10$^{22}$ cm$^{-2}$ and in which
$d$log{\fperpNX}/$d$log{$N_{\perp}$} decreases below $-$ 3.0 as
{$N_{\perp}$} increases toward $N_{0}$. This is observed locally and
is presumably due to the conversion of H I into H$_{2}$ at large
column densities (Zwaan {\etal}  2006). However, 
the molecular fraction does not increase with $N$ in DLAs
(Ledoux {\etal} 2003), which is likely due to low dust content. Since
the gamma distribution fit, {\fNX} $\propto$
$(N/N_{\gamma})^{\alpha_{2}}$exp($-N/N_{\gamma})$ (PHW05), leads to
negative {\fperpNX} for {$N_{\perp}$} $>$
(3$+{\alpha_{2}})$$N_{\gamma}$, we conclude that our double power-law
fit 
to {\fNX,
with $\alpha_{4}$ = $-$ 3.0,
is the most plausible solution for DLAs, provided they
are disk-like structures, and we adopt it here and in what follows.

\begin{figure}
\figurenum{1}
\includegraphics[angle=270,width=.80\textwidth]{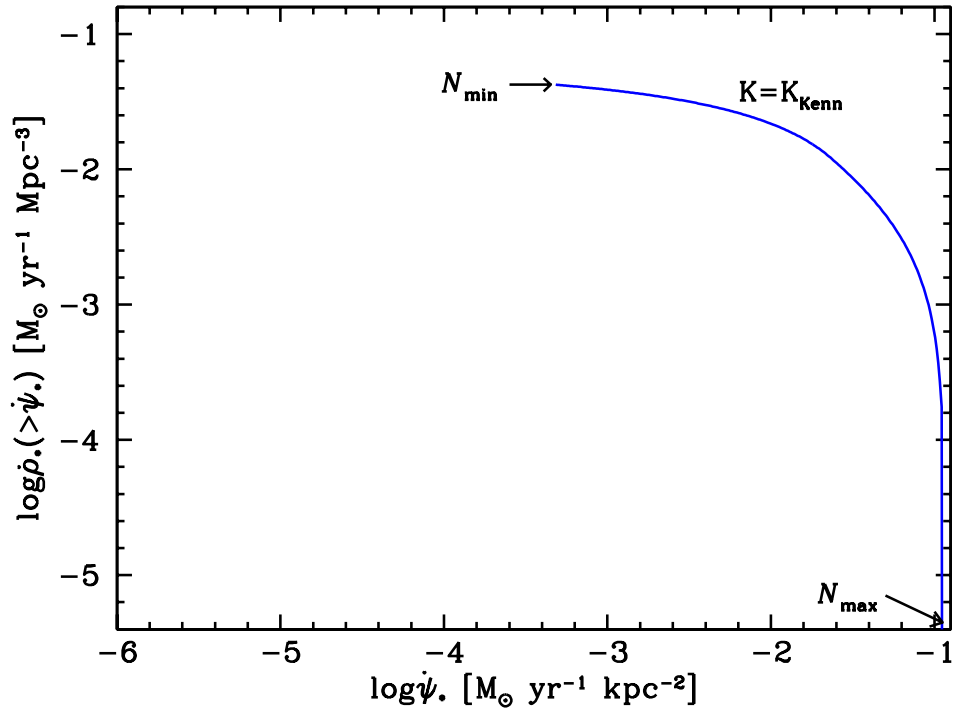}
\caption[]{Cumulative comoving SFR density 
versus {\ps}. 
Curve is {\rhodotpsi} predicted
for Kennicutt-Schmidt law with $K$=$K_{Kenn}$.
The range of {\ps} 
corresponds to $N$=[2{$\times$}10$^{20}$,1{$\times$}10$^{22}$].} 
\label{fig:rhodotth}
\end{figure}

The resulting cumulative comoving SFR density is depicted by the blue
curve in Fig.~{\ref{fig:rhodotth}}. We plot {\rhodot}(${\ge}${\ps},$X$)
for the range in {\ps} corresponding to $N=[N_{min},N_{max}]$,
where here and unless otherwise noted,
$N_{min}$=2{$\times$}10$^{20}$ cm$^{-2}$ and $N_{max}$=10$^{22}$
cm$^{-2}$. By adopting a value for  $N_{min}$ that is lower 
than the range of threshold
column densities $N_{\perp}^{crit}$ observed for nearby
galaxies (Kennicutt 1998b), we overestimate {\rhodot}($\ge$$N_{min}$).
However,  
the upper limits on {\rhodot} estimated in $\S$ 3.3
are valid only for  {\nh} 
higher than the local values of $N_{\perp}^{crit}$,
and as a result comparison between theory and observation occurs
at column densities where the Kennicutt-Schmidt law is well
established.
We also assume
$z$=3 and use Eq.~{\ref{eq:Inuav}} and the Kennicutt parameters
to convert $N$ to {\ps}. In agreement with Hopkins {\etal} (2005) we
find that {\rhodot} inferred for DLAs is predicted to be somewhat
lower than deduced for LBGs (Steidel {\etal} 1999; Giavalisco {\etal}
2004), or for DLAs from the {\ciis} technique 
(Wolfe {\etal} 2003a [hereafter WGP03]).

\subsection{Expected Number of DLAs in the UDF}
How many DLAs are expected to
populate the UDF? The answer depends on three factors: (1) the
column-density distribution function {\fNX}, (2) the redshift search
interval $z=[z_{low},z_{high}]$, and (3) the linear sizes of the DLAs.
Suppose all DLAs were equal in size. Because angular diameter is an
insensitive function of $z$ at $z$ $>$ 1.2, we assume that all DLAs in
the search volume have identical angular diameters,
{$\theta_{DLA}$}. In that case the number of DLAs with
column-densities greater than $N$ is given by

\begin{equation}
{{\cal N}({\ge}N)}={\Biggl (}{{\omega_{UDF} \over {{\pi}{\theta_{DLA}^{2}}/4}}{\Biggr )}{\int_{X({z_{low}})}^{X({z_{high}})}dX}{{\int_{N}^{N_{max}}}dN'f(N',X)}}
\label{eq:NumberDLA}
\cmma
\end{equation}

\begin{figure}
\figurenum{2}
\centering
\includegraphics[angle=0,width=.80\textwidth]{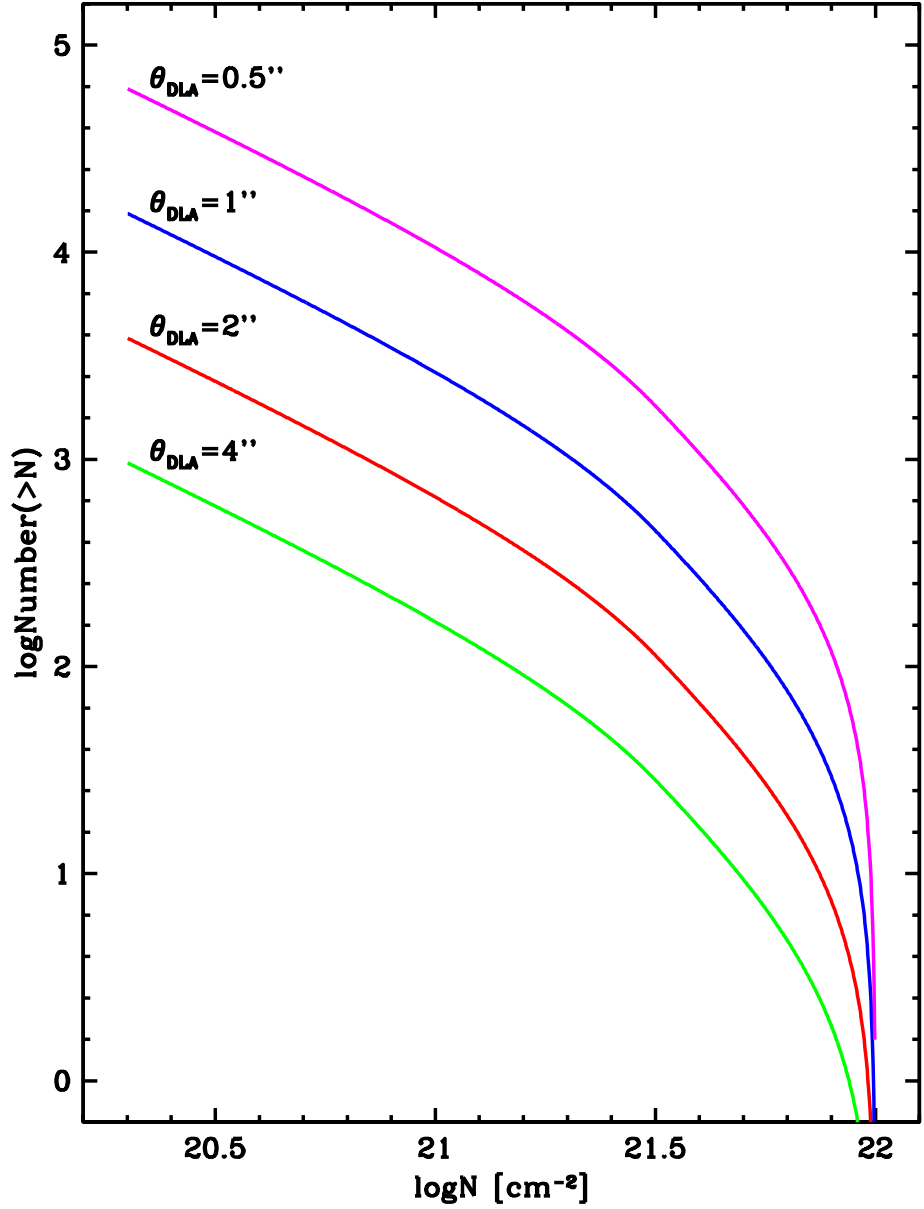}
\caption[]{Plot showing number of DLAs with column densities
exceeding $N$ predicted to populate the UDF in the
redshift interval $z$=[2.5,3.5]. Magenta, blue, red, and green curves
are for DLAs with {$\theta_{DLA}$} =0.5{\arcsec}, 1{\arcsec}, 2{\arcsec}, 
and 4 {\arcsec}.} 
\label{fig:NumberDLA}
\end{figure}

\noindent where $\omega_{UDF}$ is the solid angle subtended by the UDF.
The
results shown in Fig.~{\ref{fig:NumberDLA}} were obtained assuming
$\omega_{UDF}$=10 arcmin$^{2}$, $z_{low}$=2.5, and $z_{high}$=3.5.
The four curves were
computed for $\theta_{DLA}$=0.5{\arcsec}, 
1{\arcsec}, 2{\arcsec}, and 4{\arcsec},  
which correspond to
linear diameters of 4, 8, 15, and 31 kpc at $z$=3 that span the range of
sizes predicted for DLAs,
which can reach diameters as low as 2 kpc (e.g. Haehnelt {\etal} 1998; Nagamine {\etal} 
2005; Razoumov {\etal} 2005; Prochaska \& Wolfe 1997). 
To evaluate the integral in
Eq.~{\ref{eq:NumberDLA}} we combined the redshift evolution of $d{\cal
N}/dX$ with the $N$-dependence of {\fNX} determined from the full
sample of DLAs found in the SDSS survey (PHW05).
The figure shows the UDF should be populated by large numbers of DLAs.
In our example of $N \ > $1.6{$\times$}10$^{21}$ cm$^{-2}$ we find
that over 100 DLAs would be present with $\mu_{V}$ $<$ 28.4 mag
arcsec$^{-2}$ if $\theta_{DLA}$ = 4{\arcsec}.  Of course, not all DLAs
will have the same size, and most should have {$\theta_{DLA}$ $<$ 
4{\arcsec}, which results in an increase in ${\cal N}$. The point is that
the UDF should contain hundreds to tens of thousands of high column-density
DLAs which according to Eq.~{\ref{eq:Inuav}} should be detectable. By
comparison,
were  the LBG luminosity function (Adelberger \& 
Steidel 2000) extrapolated
to an approximate UDF point-source limit of $V$=30.5,
the UDF would be populated with 1800 LBGs in the
redshift interval $z$=[2.5,3.5].

\section{ANALYSIS}

We now describe the results of our search for emission from DLAs.

\subsection{Object Detection}

To determine {\rhodotpsi} (here and in what follows
we let {\rhodotpsi}$\equiv${\rhodot}[$\ge${\ps},$X(z=3)$])
empirically, we searched for extended, low
surface-brightness emission in the UDF.  The search was performed with
images acquired with the F606W filter. The wavelength centroid of the
F606W filter approximately matches the rest-frame FUV wavelength of
1500 {\AA} for the redshift search interval $z$=[2.5,3.5]. This is the
standard rest-frame wavelength for which the LBG luminosity function
and comoving SFR densities have been determined (Steidel {\etal} 1999;
Adelberger \& Steidel 2000).  Furthermore, this redshift interval
contains the largest sample of known LBGs, which we shall use for
comparison with our results, and the F606W image is the most sensitive
one obtained in the UDF.  Because the UDF does not have the $U$-band
sensitivity required for this project
and the Lyman limit discontinuities of $z\sim 3$ objects
occur further blueward of the blue filter response, photometric
redshifts for objects in this redshift range are subject to
catastrophic errors.  We therefore apply the data only for
determining {\em
upper limits} on {\rhodotpsi} by assuming that all objects lie in
this redshift range.

We performed object detection in the drizzle-combined F606W image
using SExtractor (Bertin \& Arnouts 1996).  The object detection
criteria are set according to the following steps.  First, we require the
presence of flux over a minimum, contiguous area comparable to the
size of an adopted kernel.  Next, we adjusted the detection threshold
to the lowest value where zero detections are found in the negative
image.  This procedure was first performed in the original F606W image
with a smoothing kernel matching the PSF of the image. Specifically,
we let the FWHM of the kernel 
{\thetkern}=$\theta_{\rm PSF}=0.09\arcsec$, in order to identify all the
high surface brightness (HSB) objects; i.e., objects with
$\mu_{V}$ $<$ 26 mag arcsec$^{-2}$.  We then masked out all these
known objects in the image by setting the associated pixels to the
median sky
value, in order to produce a masked-F606W image.  To search for
extended low surface brightness objects, we smoothed the masked-F606W
image using circular Gaussian kernels with {\thetkern}={\thetdla}
where {\thetdla} is the FWHM of the DLA image and let the
aperture diameter equal {\thetkern}. 
The advantage of smoothing is that
in the limit of random noise the SNR of the surface brightness of the
smoothed image is enhanced relative to the unsmoothed image by the 
factor {\thetkern}/{$\theta_{\rm PSF}$} (e.g. Gonzalez {\etal} 2001; Irwin
{\etal} 1985). 
We emphasize that removing HSB objects detected
in the unsmoothed image from all subsequent searches 
allows us to search for extended low surface-brightness
features closer to these HSB objects in smoothed images. 
Consequently we are testing the hypothesis that emission
from DLAs arises only from star formation in
the gas detected in absorption, since the Kennicutt-Schmidt
law predicts $\mu_{V}$ to be fainter than 26 mag arcsec$^{-2}$
for {\nh}=[$N_{\rm min},N_{\rm max}$]
and $z$ = [2.5,3.5]. 
\footnote{We chose this 
surface-brightness limit  because at $z$ =  3 it
corresponds to $N$ $<$ 8{$\times$}10$^{21}$ cm$^{-2}$, the highest
column density yet measured in the SDSS DLA survey (PHW05).} 

At $z$ $>$2 most HSB objects are LBGs. While the brighter, $V$ $<$ 25,
LBGs have half-light radii, $r_{hl}$$\approx$2kpc (Giavalisco {\etal}
1996), recent studies with the UDF indicate $r_{hl}$ is smaller for
fainter objects and that $r_{hl}$ decreases 
with increasing redshift. Specifically Bouwens {\etal} (2004) find that the
mean half-light radius, $<r_{hl}>$=0.9 kpc for their sample of 
intermediate magnitude,
$z_{850,AB}$ $<$ 27.5, objects. We shall assume a conservative upper limit 
$<r_{hl}>$$<$1.5 kpc for LBGs in the UDF with $V$ $<$ 30. 
The light profiles of LBGs are well fitted with exponentials. Since
exponentials 
have FWHM linear diameters equal to 2$r_{hl}$/2.5, the expected
FWHM angular diameters for most LBGs in our sample, 
$\theta_{LBG}$$<$0.13{\arcsec}.  By contrast the angular
diameters predicted for the light-emitting regions
of DLAs are systematically larger.
Although the 
observational evidence for redshifts exceeding 1.9 
suggests 2.0 kpc $<$ $d_{\rm dla}$ $<$ 15 kpc
(where $d_{\rm dla}$ is the FWHM linear diameter),  
the data sample contains only three objects (WGP05).
On the other hand, 
theoretical arguments suggest DLAs are larger than LBGs.
While the predicted sizes of DLAs
are smallest for CDM models, most of the predictions for  $z$ $\sim$ 3
indicate $d_{\rm dla}$ $>$ 1.9 kpc. 
Comparison with the numerical simulations of
Nagamine {\etal} (2006) shows that 70 $\%$ is 
a conservative lower limit for the fraction of $z$=3 DLAs with
$d_{\rm dla}$ $>$ 1.9 kpc. Similarly, Haehnelt {\etal}
(2000) predict at least 80 $\%$ of these DLAs to have
$d_{\rm dla}$ $>$ 1.9 kpc, while the lower limit
computed by Mo {\etal} (1998) is 95 $\%$. Because other
models (e.g. Prochaska \& Wolfe 1997; Boisser etal 2003) predict
even larger sizes, current ideas about DLAs suggest that 
the
bulk of DLAs predicted in current models have linear diameters
larger than 1.9 kpc, which corresponds to 0.25 {\arcsec} for
the redshift interval $z$=[2.5.3.5].
We shall revisit
the possible connection between DLAs and HSB objects in $\S$ 5.2.

\subsection{Search Results}

Using the {\thetkern}=0.09 {\arcsec}
smoothing kernel, we identified roughly 11,000 objects  brighter
than the survey limit of
$V\le 30.5$ and with
$\langle\mu_{V}\rangle\le 26.6$ mag arcsec$^{-2}$ 
in the original F606W image
over the central 10 arcmin$^2$ of the UDF. 
Interestingly, none of the
objects satisfied our search criteria for {\em in situ}
star formation throughout the DLA gas; i.e., 
low surface brightness,
$\mu_{V}$ $>$ 26 mag arcsec$^{-2}$,
and angular diameter exceeding 
0.25 {\arcsec}, which corresponds to the
smallest linear {\em diameter} \footnote{Most discussions about DLA sizes refer to 
impact parameter $b$ rather than diameter $d$. In the simple case
of randomly oriented circular disks, the median
impact parameter $b$=$d$/2{$\sqrt 2$}, which
corresponds to $b$ = 0.67 kpc for {\thetkern} =0.25 {\arcsec}.}
 predicted for a substantial
fraction of model DLAs ($\S$3.1).
We considered the ``tadpole galaxies''
identified at $z<4$ in
the UDF by Straughn {\etal} (2006) as potential candidates because
they typically consist of multiple compact components distributed over
an area $\approx$ 0.1{$\times$}1 arcsec$^{2}$:  
we confirm
the presence and
properties of all 163 ``tadpole galaxies''
identified by Straughn {\etal} (2006)  and measure their average
$V$ surface brightness $<{\mu_{V}}>$ = 26.6 mag arcsec$^{-2}$.
These authors detected between 50 and 60 galaxies
with provisional 
redshifts between 2.5 and 3.5, which leads to 
an area covering factor $f_{A}$
$\approx$ 10$^{-4}$. By contrast $f_{A}$=0.33 for all DLAs 
in this redshift interval, and
$f_{A}$=0.13 for DLAs with $N$ $\ge$ 5{$\times$}10$^{20}$ cm$^{-2}$,
the H I column density above which the Kennicutt--Schmidt law
is firmly established (Kennicutt 1998a,b). As a result the ``tadpole galaxies''
cannot be the sites of {\em in situ} star formation in 
most DLAs. On the other hand, the relatively high surface brightness
of the compact components in these galaxies 
is consistent with star formation occurring
in regions with $N$ $>$ 5{$\times$}10$^{21}$ cm$^{-2}$ that follow
the Kennicutt--Schmidt law.

While we have not attempted to measure the redshifts of
the remaining high surface-brightness sources in our sample,
the majority of objects with
$\mu_{V}$ $<$ 26.0 mag arcsec$^{-2}$ 
are likely to be 
faint blue galaxies with $z$ $<$ 1.5.
This conclusion is based on the luminosity function 
presented by Blanton {\etal} (2005), which includes a double
Schechter function to account for the steep rise in galaxy
density at the faint end. At $r'$ = 28 about 10 $\%$ will
be at $z$ $>$ 3, with a median redshift of $z$ = 1, while
at $r'$=30 about 30$\%$ will be at $z$ $>$3 with a median
redshift of 1.5. These numbers are relevant to the
F606W image because its
wavelength centroid is similar to that of the $r'$ filter. 


 To increase the probability for detecting extended low surface
brightness objects, we searched the masked F606W
image by convolving it with smoothing kernels matched to a range of predicted
{\thetdla}'s, from {\thetdla}=0.25{\arcsec} to {\thetdla}=4.0{\arcsec} (see
\S\ 2.2).
The number of objects found in the smoothed images versus the
smoothing kernel size is presented in Fig.~{\ref{fig:dlanumvthet}}.
The figure shows that the number of detected objects decreases rapidly
with increasing {\thetkern}: while 166 objects are detected when
{\thetkern}=0.25{\arcsec}, only one object is found when
{\thetkern}=1.0{\arcsec}, and {\em no
objects are found for {\thetkern} $>$ 1.0{\arcsec}.}
In $\S$ 3.3 we shall argue that the steep fall off with {\thetkern}
indicates the presence of compact sources
with light profiles that decline steeply with radius. 

\begin{figure}
\figurenum{3}
\includegraphics[angle=270,width=.80\textwidth]{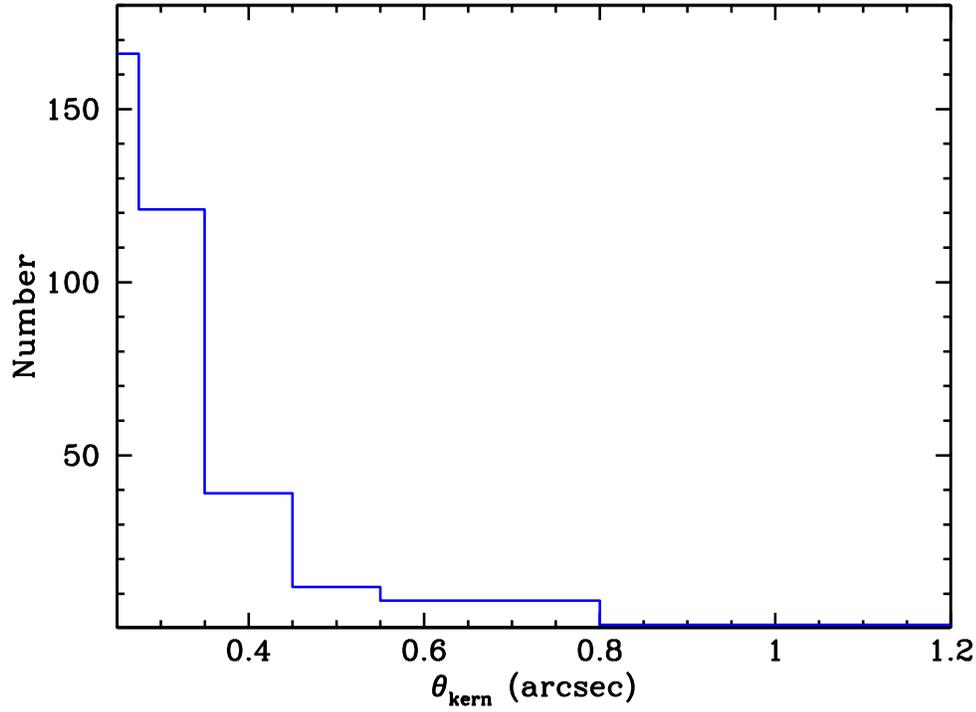}
\caption[]{Histogram showing number of recovered DLAs as
a function of {\thetkern}. Steep decrease with increasing {\thetkern}
indicates that increase in aperture size results in increase
in sky noise but not in signal. This implies the predominance of compact
galaxy light profiles reminiscent of dwarf galaxies at intermediate
redshift. Part of fall-off is also due to the observed intrinsic decrease
in galaxy counts with increasing flux. 
}
\label{fig:dlanumvthet}
\end{figure}

The implication is that spatially extended
sources of low surface-brightness emission are rare in the UDF.
Fig. 4 displays multi-color images of the 12 objects
detected for {\thetkern}=0.5{\arcsec}.
Comparison between the
smoothed and unsmoothed F606W image demonstrates the efficiency of the
matched kernel technique for detecting extended objects that are
exceedingly faint.  
The detection of objects with
{\thetkern} = 0.5{\arcsec} supports the conclusion that had they been
present, faint objects with {\thetdla} {$>$} 1{\arcsec} would also have
been detected with kernels appropriately matched to {\thetdla}.
Lacking photometric
redshifts, one cannot determine the nature of the 12 objects but it is
likely that they are low surface-brightness
dwarf galaxies at redshifts $z\sim 1$. The blue colors,
central surface brightnesses, and linear
diameters of the sample compiled
by van Zee {\etal} (1997) would give rise to $V$ band magnitudes,
surface brightnesses, and angular diameters similar to those detected here,
if the sample were redshifted to $z$ $\approx$ 1.
Because of the similarity in $V$ magnitudes,
we also considered whether our detected objects belong to the class of IRAC
selected galaxies found in the UDF by Yan {\etal} (2002).
But this is unlikely since the IRAC galaxies are brighter
at $z_{850,AB}$ magnitude than at $V$, whereas all of the  objects
that we detected in $V$  with {\thetkern}=0.25 {\arcsec}
to 1.0 {\arcsec} are either fainter or undetected
in $z_{850,AB}$. 
As a result, the density
obtained by assuming
these objects are in the redshift interval $z$=[2.5,3.5] acts as a
conservative upper limit to the density of DLAs with
{\thetdla}=0.5{\arcsec} in the same redshift interval.
Similar conclusions hold for all DLAs with {\thetdla}
${\le}$ 1{\arcsec}.

\begin{figure}
\figurenum{4a}
\caption[]{multi-band images of galaxies detected with
0.5 {\arcsec} kernel. Column 1  shows smoothed F606 W image, col. 2
shows F606W masked image. Remaining columns show unsmoothed images
for designated filters. Images are 5 {\arcsec} on each side.}
\label{fig:0p5secimage}
\end{figure}

\begin{figure}
\figurenum{4b}
\caption[]{Same as 4a}
\label{fig:0p5bsecimage}
\end{figure}

\begin{figure}
\figurenum{4c}
\caption[]{Same as 4a}
\label{fig:0p5csecimage}
\end{figure}

Therefore, in the cases of null detections, {\thetdla} $>$ 1 {\arcsec},
we use cumulative
Poisson probabilities
to place a 95\% confidence upper limit on {$\cal N$} of 
{$\cal N_{\rm 95}$} = 3 for the
number of low surface-brightness DLAs 
populating the UDF in the redshift interval $z$ = [2.5,3.5].
In the case of a single detection, {\thetdla} = 1 {\arcsec},
we place a corresponding
95 $\%$ confidence upper limit of {$\cal N_{\rm 95}$} = 4.7. 
For the multiple detections at {\thetdla}
$<$ 1 {\arcsec}, the values of   {$\cal N_{\rm 95}$}
are given in Table 1.
We emphasize that the upper
limits on {$\cal N$} are valid only for objects brighter than some
threshold magnitude and surface brightness, which we determined by a
Monte-Carlo technique described in the following subsection  
($\S$ 3.3).

\begin{table} 
\begin{center}
\begin{tabular}{lcccccccc}
{\thetexp}$^{a}$&{\thetkern}$^{b}$&{$\cal N_{\rm det}$$^{c}$}&{$\cal N_{\rm 95}$}$^{d}$&$V^{thresh}$ \ $^{e}$&${\mu^{thresh}_{V}}$ \ $^{f}$&SFR$^{g}$&log({\ps})$_{thresh}$$^{h}$&log{\rhodot}$^{i}$\\
arcsec &arcsec& &&mag&mag arcsec$^{-2}$&{\smpy}&{\smpykpc}&{\smpympc}\\
(1)&(2)&(3)&(4)&(5)&(6)&(7)&(8)&(9) \\
\tableline
0.18 &0.25&166&$<$186&....&28.0 &14.0&$-$1.73&$-$3.31  \\
0.22 &0.3&121&$<$141&....&28.1 &13.5&$-$1.77&$-$3.32\\
0.29 &0.4&39&$<$50.9&....&28.5 &7.15&$-$1.93&$-$3.59  \\
0.36 &0.5&12&$<$19.4&....&28.7 &2.91&$-$2.01&$-$3.84  \\
0.43 &0.6&8&$<$14.4&....&28.9 &2.76&$-$2.09&$-$3.81  \\
0.72 &1.0&1&$<$4.7& ....&29.3 &0.66&$-$2.25&$-$4.09  \\
1.44 &2.0&0&$<$3&26.5&29.8 &2.86&$-$2.45&$-$3.58  \\
2.89 &4.0&0&$<$3&25.0&29.7 &11.7&$-$2.41&$-$2.97  \\
\end{tabular}
\end{center}
\caption{Results of UDF Search in F606W Image} \label{data}
\tablenotetext{a}{Light profile exponential scale-length for synthetic
galaxy used to compute threshold value of {\ps} for all cases
and threshold {\rhodot} in the case of null detections.}
\tablenotetext{b}{FWHM of Gaussian smoothing kernel}
\tablenotetext{c}{Number of objects detected  in the UDF with
given value of {\thetkern}}
\tablenotetext{d}{95 $\%$ confidence upper limit on number of objects
in UDF with given {\thetkern}}
\tablenotetext{e}{Threshold input  $V$ magnitude,
when $\epsilon$=0.2. Used to compute upper limits on 
SFRs for kernels with null detections. Not used for kernels
with positive detections.}
\tablenotetext{f}{Threshold input
$V$ band surface brightness when $\epsilon$=0.2}
\tablenotetext{g}{Upper limit on SFR.
For {$\cal N_{\rm det}$} = 0 this corresponds to threshold $V$ magnitude,
and for 
{$\cal N_{\rm det}$} $\ne$ 0 this corresponds the SFR averaged over
the {$\cal N_{\rm det}$} galaxies.}
\tablenotetext{h}{Lower limit on {\ps}($z$=3.5) corresponding to threshold ${\mu_{V}}$}
\tablenotetext{i}{95 $\%$ confidence upper limit on {\rhodot}}
\end{table}

\subsection{Search Completeness}

To determine the sensitivity of our searches for extended low 
surface-brightness objects, we performed a series of Monte-Carlo simulations.
First, we generated test objects of different brightnesses for
exponential light profiles of different scale-lengths.  The angular exponential
scale-lengths $r_s$ were adjusted such that the FWHM
({$\equiv$}[$2{\ln\,2}$]{\thetexp}) = {\thetdla}.
Next, we placed the
test objects of a given magnitude and scale-length at 1000 random
positions 
in the masked-F606W image, and smoothed the images with
kernels in which {\thetkern} = {\thetdla}
in the optimal case of matched kernels. 
Finally, we
performed the detection procedure following the same detection
algorithm described in \S\ 3.1 for finding low surface brightness
objects.  We also repeated the simulations in which the synthetic
galaxy light profiles were elliptical exponentials with $b/a$=0.5
and with position angles varying between 0$^{o}$ and 90$^{o}$:
the difference with results for circular apertures was
insignificant.

\begin{figure}
\figurenum{5}
\includegraphics[angle=270,width=.80\textwidth]{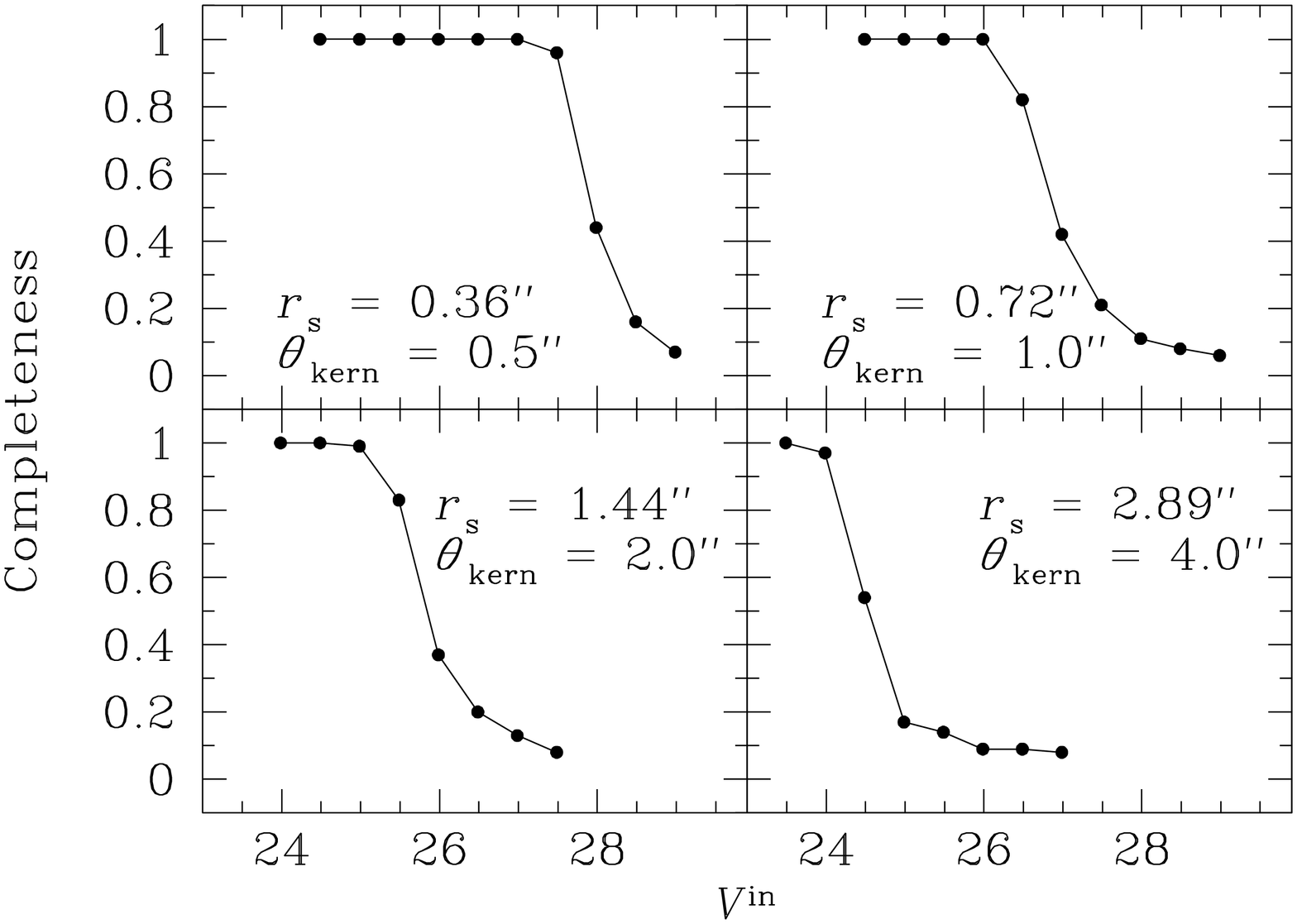}
\caption[]{Recovery fraction $\epsilon$  versus input magnitude $V^{in}$
for increasing values of {\thetkern} for the matched kernel
condition, {\thetkern}={\thetdla}. Notice how $V^{in}$ at a given
recovery fraction, $\epsilon$=0.2, decreases with increasing
{\thetkern}. This results from noise increase due to increase
in aperture size. The flattening observed at $\epsilon$=0.2
is due overlapping of synthetic galaxy profiles with low brightness
wings extending from masked galaxies.}
\label{fig:efficiency}
\end{figure}

We measured the photometric properties of the recovered galaxies and
computed the recovery fraction, $\epsilon$. This procedure was
repeated by increasing the input magnitude $V^{in}$ until the
recovered number of galaxies decreased to some critical value. The
results of this procedure are illustrated in
Fig.~{\ref{fig:efficiency}}, which plots $\epsilon$ versus $V^{in}$.
In
principle one could determine the threshold value of $V$ by raising
$V^{in}$ until only {$\cal N_{\rm 95}$}  objects are left in the image.
For those cases where {$\cal N_{\rm 95}$} $<<$ 200 (i.e. $\epsilon$
$<<$ 0.2) we found that the measured magnitudes, $V^{meas}$, were sufficiently
faint that overlap between the synthetic galaxies and the low-surface brightness
wings extending from the masked real galaxies led to spurious over-estimates
in flux determinations for the synthetic galaxies
(in most cases this effect results in $V^{meas}$ to be brighter
than $V^{in}$).  This is evident
in the saturation of the $\epsilon$ versus $V^{in }$ curves
in Fig.~{\ref{fig:efficiency}}: the flattening at $\epsilon$ $<$ 0.2
for {\thetdla}=4 {\arcsec} and at $\epsilon$
$<$ 0.1 for the other values of {\thetdla} illustrates this effect.
On the other hand when $\epsilon$ $\ge$ 0.2, the $V^{in}$ were
sufficiently bright that the output $V^{meas}$ were reliable. As
a result we assume the threshold magnitude is given by
$V^{in}$ for $\epsilon$ = 0.2 and the threshold surface brightness
is given by the corresponding value of $\mu_{V}^{in}$
(hereafter we refer to these as $V^{thresh}$ and
$\mu_{V}^{thresh}$). We emphasize
that the
$\epsilon$=0.2 criterion
is conservative as it requires the detection 
of 200 objects, which exceeds {$\cal N_{\rm 95}$} 
for all values of {\thetdla}.
The resulting threshold magnitudes and surface
brightnesses corresponding to the values of
{\thetkern} used for galaxy detection 
are shown in Table 1.

A possible concern over this procedure arises from
our use of 10$^{3}$ synthetic galaxies even though
Fig.~{\ref{fig:NumberDLA}} shows
that {${\cal N}$}, 
the number of DLAs
predicted for the UDF, can be as large as 10$^{5}$.
However, 
the threshold values of $V^{in}$
and $\mu_{V}^{in}$ inferred from the simulations are
independent of {${\cal N}$}. 
The reason is that confusion noise due to fluctuations in the number of
synthetic galaxies per aperture beam does not increase
with increasing {$\cal N$}: for DLAs with a fixed
area covering factor $f_{A}$(DLA) the  
number of
galaxies per ``beam''  area, $\pi${\thetkern}$^{2}$/4, 
equals  $f_{A}(DLA)$({\thetkern}/{\thetdla})$^{2}$, which
is independent of {${\cal N}$} and
is less than 1 for the case of matched kernels assumed here.
Other systematic errors such as that 
due to overlap between the
galaxy light profiles is negligible because
the area covering factor of 200 synthetic galaxies is less
than 6 $\%$. Therefore, the principal systematic error 
in determining the thresholds is caused by
overlap between the synthetic light profiles and the faint wings
of masked real galaxies (see above). 


\begin{figure}
\figurenum{6}
\includegraphics[angle=0,width=.80\textwidth]{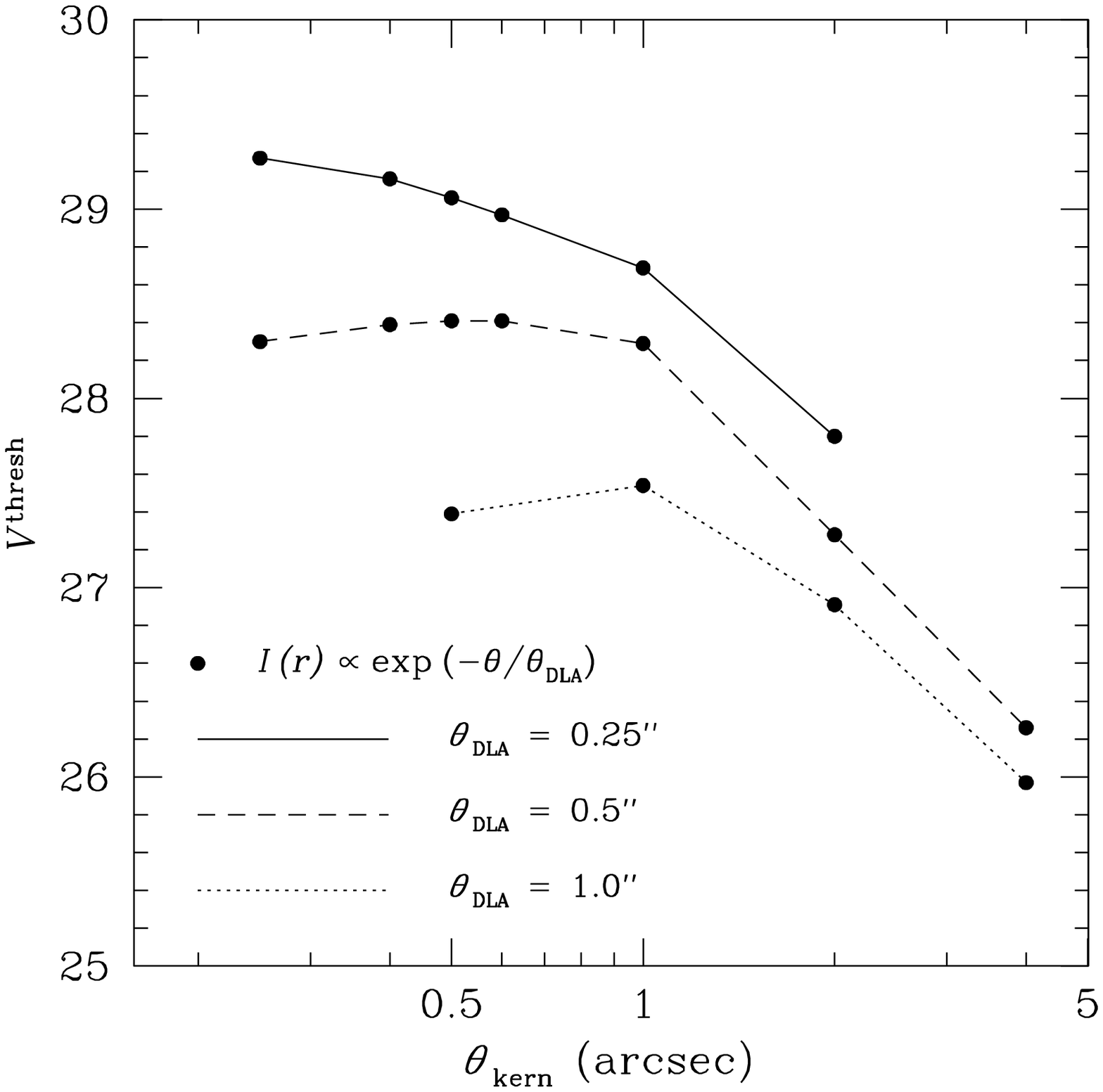}
\caption[]{Threshold magnitude $V^{thresh}$ versus {\thetkern} for fixed 
DLA diameters, {\thetdla} = 0.25{\arcsec}, 0.5{\arcsec}, and 1.0 {\arcsec}.
Decrease of $V^{thresh}$ with increasing {\thetkern} at {\thetkern} $\ge$
{\thetdla} due to increase in sky noise in larger apertures. Decrease
of $V^{thresh}$ with decrease in {\thetkern} at {\thetkern} $<$ {\thetdla}
due to effect of smoothing kernel alone. }
\label{fig:bestmag_april13}
\end{figure}

We next used the simulations to interpret the
steep fall-off in the number of detected galaxies with {\thetkern}
(Fig.~{\ref{fig:dlanumvthet}}). Part of the fall-off
is due to the loss of sensitivity resulting from the
increase of {\thetkern} above {\thetdla}. This is
illustrated in Fig.~{\ref{fig:bestmag_april13}}, which
shows that
$V^{thresh}$ decreases with increasing {\thetkern}
for {\thetkern} $>$ {\thetdla}.
Therefore, a loss of sources will result from
a decrease in survey depth with increasing {\thetkern}.
A fall off of sources with decreasing $V^{thresh}$ can
also be attributed to the intrinsic shape of the source
counts-magnitude relation.
From known field number counts (Metcalfe et al.\ 20001), 
we estimate that for every 
half magnitude deeper, we gain about 20$\%$ more objects at $V$ $\sim$28.  
If we 
adopt this rate and our survey depth estimated from
an exponential profile, which went from $V^{thresh}$=29.0 to $V^{thresh}$=28.7 
when we increase the smoothing kernel size 
from 0.3{\arcsec} to 0.4{\arcsec}, 
we estimate a loss of only about 13 $\%$ of 
the targets if the decline is due to survey depth alone. 
Because Fig.~{\ref{fig:dlanumvthet}} indicates a more significant 
drop than this, we suggest within one exponential
scale length the 
intrinsic light profiles of the detected galaxies are steeper 
than an exponential function, which is consistent
with the dwarf galaxy hypothesis. 

\subsection{Estimated Comoving SFR Densities}

\begin{figure}
\figurenum{7}
\includegraphics[angle=270,width=.80\textwidth]{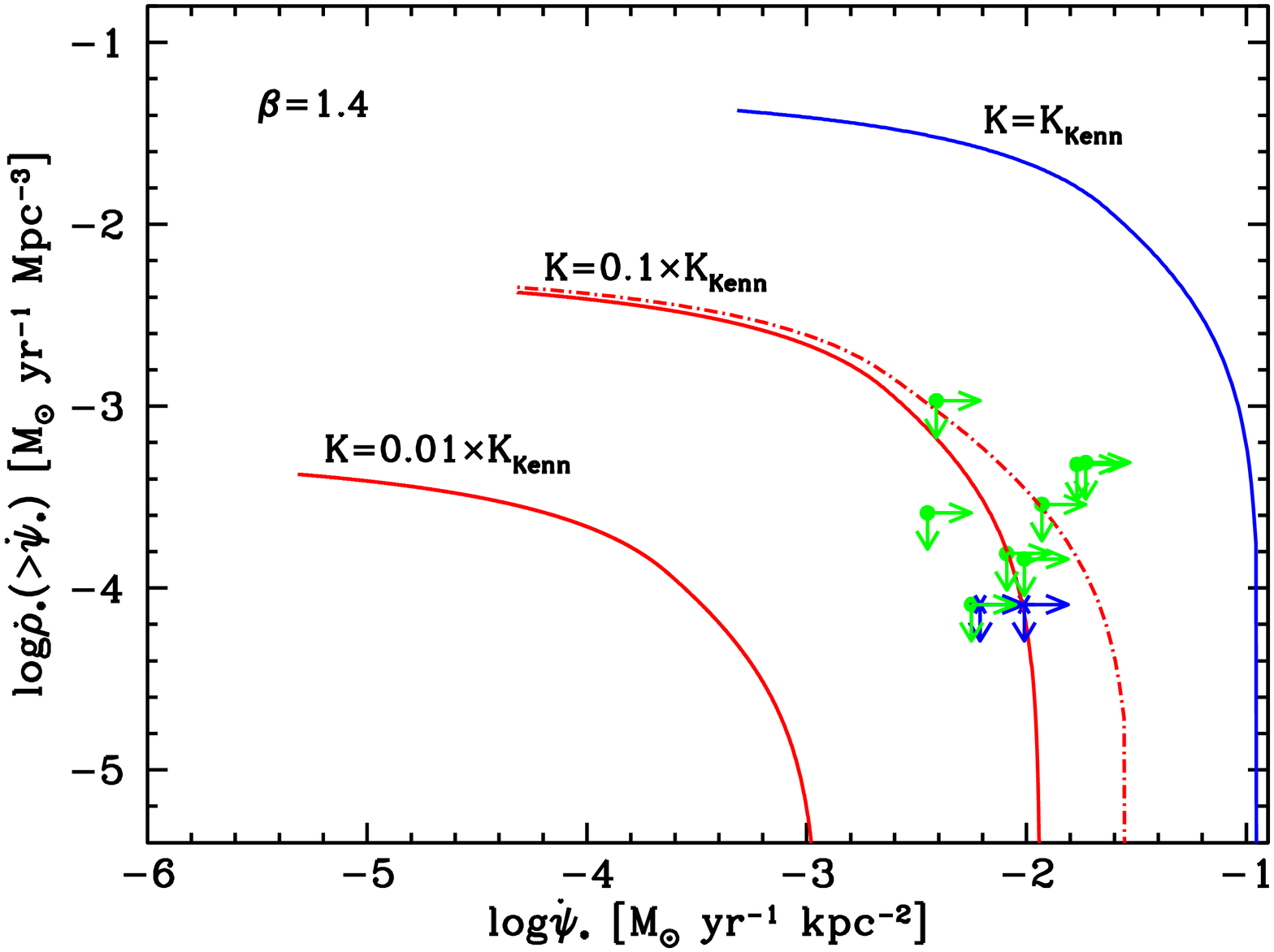}
\caption[]{Cumulative comoving SFR densities 
versus {\ps}. Green data points correspond
to upper limits derived for {\thetkern}=4{\arcsec}, 2{\arcsec}, 1{\arcsec}, 
0.6{\arcsec}, 0.5{\arcsec}, 0.4{\arcsec}, 0.3{\arcsec},
and 0.25 {\arcsec} in counter-clockwise  
order starting with (log{\ps},log{\rhodot})=($-2.41,-2.97$): 
matched kernels assumed
in deriving SFR thresholds. Blue (starred) data points
correspond to unmatched kernel with {\thetdla}=1.0 {\arcsec}
and {\thetkern} = 0.5{\arcsec} and 2.0{\arcsec} in order of
increasing {\ps}. 
Blue curve is {\rhodotpsi} predicted
for Kennicutt-Schmidt law with $K$=$K_{Kenn}$.
Solid Red curves are for
$K$=0.1$K_{Kenn}$ and $K$=0.01$K_{Kenn}$. 
The range of {\ps} for all solid curves
corresponds to $N$=[2{$\times$}10$^{20}$,1{$\times$}10$^{22}$]. 
Dot-dashed red curve is same as solid red curve for
$K$=0.1$K_{Kenn}$ except that $N_{max}$=2{$\times$}10$^{22}$ cm$^{-2}$.} 
\label{fig:rhodotvN}
\end{figure}

We computed empirical upper limits on {\rhodotpsi} in the following
manner. First, in the case 
of null detections, {\thetkern}=2{\arcsec} and 4{\arcsec} , 
we
used the threshold magnitudes  $V^{thresh}$ 
in Table 1 to determine the
minimum luminosity, $L^{min}_{\nu}$ detectable out to
$z_{high}$=3.5. We then converted $L^{min}_{\nu}$ into a SFR by
assuming SFR=1.25{$\times$}10$^{-28}L^{min}_{\nu}$ (Madau {\etal}
1998), and let
{\rhodot}={$\cal N_{\rm 95}$}{$\times$}{SFR}/{$\Delta V_{co}$} 
to find the 95\%
confidence upper limit on {\rhodot}, where {$\Delta V_{co}$}, the
comoving volume of the UDF, equals 3.3{$\times$}10$^{4}$ Mpc$^{3}$ for
the redshift interval $z$=[2.5,3.5].  We then used
Eq.~{\ref{eq:Inuav}} to compute the threshold value of {\ps} from the
threshold surface brightnesses $\mu_{V}^{thresh}$ in Table 1. The resulting {\rhodotpsi}
are plotted in Fig.~{\ref{fig:rhodotvN}} and entered in Table 2 for
both values of {\thetkern}. 
In the case of  positive detections, {\thetkern} $<$ 1{\arcsec},  
we assumed all the
detected objects were at $z$=3 (since the average redshift of 
all of the detected objects is more likely to
equal the average redshift  rather than
the maximum redshift, $z_{high}$, of the UDF search volume) and added the inferred values of
$L_{\nu}$ to obtain the total SFR. 
Because
{$\cal N_{\rm 95}$} exceeds the number of detected
galaxies, {$\cal N_{\rm det}$},  the luminosities of the
${\cal N_{\rm 95}} - {\cal N_{\rm det}}$ 
galaxies are unknown. We addressed this issue by
assigning the mean SFR of the detected sample $<$SFR$>$
to each excess galaxy. Therefore, in these cases 
{\rhodot}=
{$ {\cal N_{\rm 95}}$}$<$SFR$>$/{$\Delta V_{co}$}.
Note, whereas the
threshold values of {\ps} were computed with the
same method used in the case of null detections,
in this case the limits on {\rhodot} depend on the $V$ magnitudes measured for
each detected galaxy rather than on the value of $V^{thresh}$ used in
the case of null detections.

We wish to comment on two more points about
Fig.~{\ref{fig:rhodotvN}}. First, the horizontal green
arrows directed toward increasing {\ps}
emphasize that each upper limit on {\rhodotpsi} is valid
for projected SFRs {$\ge$} {\ps}. Second, the upper limits
on {\rhodot} decrease with decreasing {\thetkern} between
{\thetkern} = 4.0 {\arcsec} and 1.0 {\arcsec} because
{$\cal N_{\rm 95}$} is essentially unchanged while the sky noise contribution
to the measured flux
is diminished in the smaller kernels. This is reversed to
an increase in {\rhodot} with decreasing {\thetkern}
for {\thetkern} $<$ 1.0 {\arcsec} because the rapid
rise in {$\cal N_{\rm 95}$} with decreasing {\thetkern}
(Fig.{\ref{fig:dlanumvthet}}) offsets the reduction in
sky noise. The corresponding {\ps} coordinate increases
with decreasing {\thetkern} in every case due to the
increased error in surface brightness with decreasing
{\thetkern}.


\section{DISCUSSION OF RESULTS}

The most striking feature of Fig.~{\ref{fig:rhodotvN}} is that the
comoving SFR densities predicted by the local Kennicutt-Schmidt law
(the blue curve) exceed the 95 $\%$ confidence upper limits
established for the UDF by at least a factor
of 30 for {\thetkern} = 0.25 {\arcsec}, 0.30 {\arcsec}, and 4.0 {\arcsec},
and by more than a factor of 100 for other
values of {\thetkern}. 

We
believe these discrepancies are real for the following reasons.

(1) the
discrepancy  is established for a wide range of {\thetkern} values,  
indicating the results are not overly sensitive to the values of
{\thetdla}.

(2) The upper limits are likely to be
conservative; i.e., the locations of the data points are likely to
occur at values of {\ps} and {\rhodot} lower than in
Fig.~{\ref{fig:rhodotvN}}. 
In the case of null detections,
{\thetkern} = 2.0 {\arcsec} and 4.0 {\arcsec}, the
location of the upper limits in the {\ps}, {\rhodot}
plane were determined by the threshold surface brightnesses and
magnitudes determined in our simulations. In $\S$ 3.3 we
discussed why our recovery criterion of 200 objects led to
large values of {\ps} and {\rhodot}. In the case of
positive detections, {\thetkern} $\le$ 1.0 {\arcsec}, the
threshold values of {\ps} are again determined by
the simulations, but the upper limits on {\rhodot} are
set by the magnitude determinations of the detected objects.
Because a significant fraction of these objects are likely to
be intermediate-redshift dwarf galaxies, the contribution
of all the detections
to the total SFR in $\Delta V_{co}$ is a conservative upper limit.

(3) The discrepancy is not overly sensitive to
the matched-kernel assumption described in
$\S$ 3. The two blue stars in
Fig.~{\ref{fig:dlanumvthet}} correspond to 
{\ps} and {\rhodot} inferred
for {\thetdla}=1 {\arcsec} and {\thetkern} = 0.5 {\arcsec}
and 2.0 {\arcsec} . The small changes in threshold
{\ps}
support our conclusion that the discrepancy
between empirical and predicted estimates in
the {\ps}, {\rhodot} plane  
is robust. 

(4) Each of the upper limits on {\rhodot} in
(Fig.~{\ref{fig:rhodotvN}}) applies to 
DLAs of the same size. But in a realistic scenario,
DLAs contributing
to {\rhodot} would span some range in size (or in mass
since size is correlated with dark-matter mass in most
models). 
Springel \& Hernquist (2003) found that 
{\rhodot}={\omgm}{$\rho_{crit}$}$\int_{0}^{\infty}S(M,z)dM$
where $\rho_{crit}$ is the current
critical cosmological density and the multiplicity function $S(M,z)$ is the product of the
comoving density of dark-matter halos and the SFR per unit
halo mass. Therefore, the SFR density predicted for a given
smoothing kernel is given by ({\rhodot})$_{\theta_{\rm kern}}$=
[$\int_{M_{-}}^{M_{+}}S(M,z)dM/\int_{0}^{\infty}S(M,z)dM$]{\rhodot}
where $M_{-}$ and $M_{+}$ are the mass limits corresponding
to the range in {\thetdla} detectable with a given {\thetkern}.
As a result the only relevant upper limits are those corresponding
to kernels sensitive to masses contributing the bulk of the
predicted {\rhodot}; i.e., kernels for which ({\rhodot})$_{\theta_{\rm kern}}$
is not much less than {\rhodot}. Our simulations indicate that
the smoothing kernels maintain maximal sensitivity for DLAs with
{\thetdla}=(0.5 {$\rightarrow$} 1.5){\thetkern}. Thus, the kernel
with {\thetkern}=2{\arcsec} is sensitive to $z$=3 DLAs with
$d_{dla}$=8kpc to 24 kpc. In the spherical collapse model the corresponding
range in dark-matter masses is 10$^{11}${\msolar} to 10$^{13}${\msolar}
if we assume the virial radius $r_{200}$ $\approx$ 10($d_{dla}$/2) (see
$\S$ 5.1). Fig. 9 in Springel \& Hernquist indicates that
halos in this mass range contribute about 50 $\%$ of the predicted
{\rhodot} at $z$=3. This model would be ruled out by our data since
the predicted {\rhodot} is 100 times larger than the empirical
upper limit for {\thetkern}=2{\arcsec}. While
more rigorous methods are needed to verify this result, our
preliminary analysis 
indicates that the results presented here are valid in the case of distributed
DLA sizes.


(5) The values of {\rhodotpsi}
predicted for the local Kennicutt parameters, $K$ and $\beta$, are
likely to be higher than depicted by the blue curve in
Fig.~{\ref{fig:rhodotvN}}. While the predicted {\rhodotpsi} is well
determined by the accurately measured parameters such as the slope
{$\alpha_{3}$}, normalization $k_{3}$, and turn-over column density
$N_{d}$, it also depends on the uncertain value of $N_{max}$.  We
chose $N_{max}$=1{$\times$}10$^{22}$ cm$^{-2}$ because the largest
measured column density in the SDSS sample is $N$=8{$\times$}10$^{21}$
cm$^{-2}$ (PHW05). The blue curve is a lower limit because
{\rhodotpsi} would increase if $N_{max}$ were raised above
1{$\times$}$10^{22}$ cm$^{-2}$, which cannot be ruled out.
Furthermore the predicted {\rhodotpsi} would also increase if DLAs are not the planar
objects assumed in $\S$ 2,

(6) A potential problem with our results is that while the UDF measurements
sample scales exceeding 1 kpc,
the expression for {\rhodot} in Eq.~{\ref{eq:rhostardotN}} depends on
{\fNX} which is based on absorption-line measurements that sample
scales  $\sim$ 1 pc (Lanzetta {\etal} 2002). The difficulty is that 
the Kennicutt-Schmidt law has 
been established only on scales exceeding 0.3 kpc (Kennicutt {\etal} 2005).
However,
because {\fNX} typically depends on over 50 measurements per
column-density bin of width $\Delta$log{\nh}=0.1 (PHW05),
Eq.~{\ref{eq:gdN}} implies that  the corresponding
comoving SFR density $\Delta${\rhodot} is a statistical average over the
differential area ${\Delta}A$$\approx$($H_{0}/c$){\fNX}$\Delta N$/$n_{co}$
(assuming spherical symmetry). We find that
$\Delta A$ $\approx$ 0.5($N_{d}/N$)({\thetdla}/1 arcsec)$^{2}$ kpc$^{2}$.
As a result $\Delta A$ $\approx$ 2 kpc$^{2}$ for 
{\thetdla}=1 {\arcsec} at
the median 
DLA column density {\nh}=8{$\times$}10$^{20}$ cm$^{-2}$. 
Since our calculation corresponds 
to the cumulative quantity {\rhodot}($>${\nh}), 
the effective statistical area probed
by our expression will exceed a few kpc$^{2}$. Therefore,
while individual 
measurements sample scales less than 1 pc, {\rhodot} is statistically
established on scales for which the Kennicutt-Schmidt law has been established
at $z$=0.

\section{IMPLICATIONS}

The discrepancies with predictions by the Kennicutt-Schmidt
law have implications 
for the efficiency of star formation in DLAs. They 
also have broader implications concerning 
metal production and energy balance in DLAs and their relationship to 
LBGs. We discuss each
of these topics in turn.

\subsection{Star Formation Efficiency in DLAs}

The discrepancy between
predicted and measured values of {\rhodot} suggests that star
formation is less efficient in DLAs; i.e., for a given $N$, {\ps} in
DLAs is lower than in nearby galaxies. This point is illustrated by
the solid red curves in Fig.~{\ref{fig:rhodotvN}}, for which the
Kennicutt parameter $K$ is reduced by factors of 10 and 100 below the
local value, $K$=$K_{\rm Kenn}$. As
$K$ decreases, the solutions 
shift toward decreasing {\ps} and {\rhodot} along a 45$^{o}$
diagonal in the ({log{\ps}, log{\rhodot}) plane as both
{\ps} and {\rhodot}  are proportional to $K$. Because the upper limits
on {\rhodot} apply only to projected SFRs above
the  {\ps} thresholds, 
Fig.~{\ref{fig:rhodotvN}} shows that 
$K$  need not be reduced by much more than
by a factor of 10  to obtain
agreement between theory and
all the data points. 
Further reductions are unnecessary since they shift the
predicted {\rhodotpsi} curves
below the {\ps} thresholds.  
Nevertheless $K$ must be reduced by {\em at least} a factor of 10.
This is illustrated by the dashed red curve in
Fig.~{\ref{fig:rhodotvN}}, which shows that {\rhodotpsi} predicted for
$K=0.1{\times}K_{Kenn}$ is incompatible  with all the upper limits
other than those
corresponding to 
{\thetkern} =0.25 {\arcsec} and  0.30 {\arcsec} when $N_{max}$ 
is increased to
$\ 2{\times}10^{22}$ cm$^{-2}$. 

The SFR efficiencies can also be decreased by lowering the
slope {$\beta$} of the Kennicutt-Schmidt law,
while keeping $K$=$K_{Kenn}$.
Although Kennicutt (1998a) found {$\beta$}=1.40$\pm$0.15 when
he compared projected SFRs with the total gas column
densities for a sample of 60 galaxies,
Gao
\& Solomon (2004) found {$\beta$}=1 by comparing
the SFRs with the H$_{2}$ column densities of 32 galaxies.
Because
our knowledge of physical
conditions in high-$z$ gas is limited, we cannot rule out the
possibility that {$\beta$} = 1 also holds for the atomic gas
content of DLAs. Fig.~{\ref{fig:rhodotvN_bet}} compares the 
{\rhodotpsi} curves predicted for {$\beta$} = 1 
and {$\beta$} = 0.6
with the standard case, {$\beta$}=1.4. With the possible 
exception of DLAs with {\thetdla}=0.25 {\arcsec}, the
curves predicted for {$\beta$}=1 are inconsistent with the data. 
Fig.~{\ref{fig:rhodotvN_bet}} also shows consistency
with all the data points is achieved if {$\beta$} $\le$ 0.6.
While we cannot rule out such low values entirely,
the cases $\beta$=1.0 or 1.4 are at least 
physically motivated (see Elmegreen 2002 and Kravtsov 2003).
Because we are unaware of any evidence or physical argument to 
justify values of $\beta$ less than 0.6, we focus
instead on other mechanisms for reducing SFR efficiencies
in DLAs.

\begin{figure}
\figurenum{8}
\includegraphics[angle=270,width=.80\textwidth]{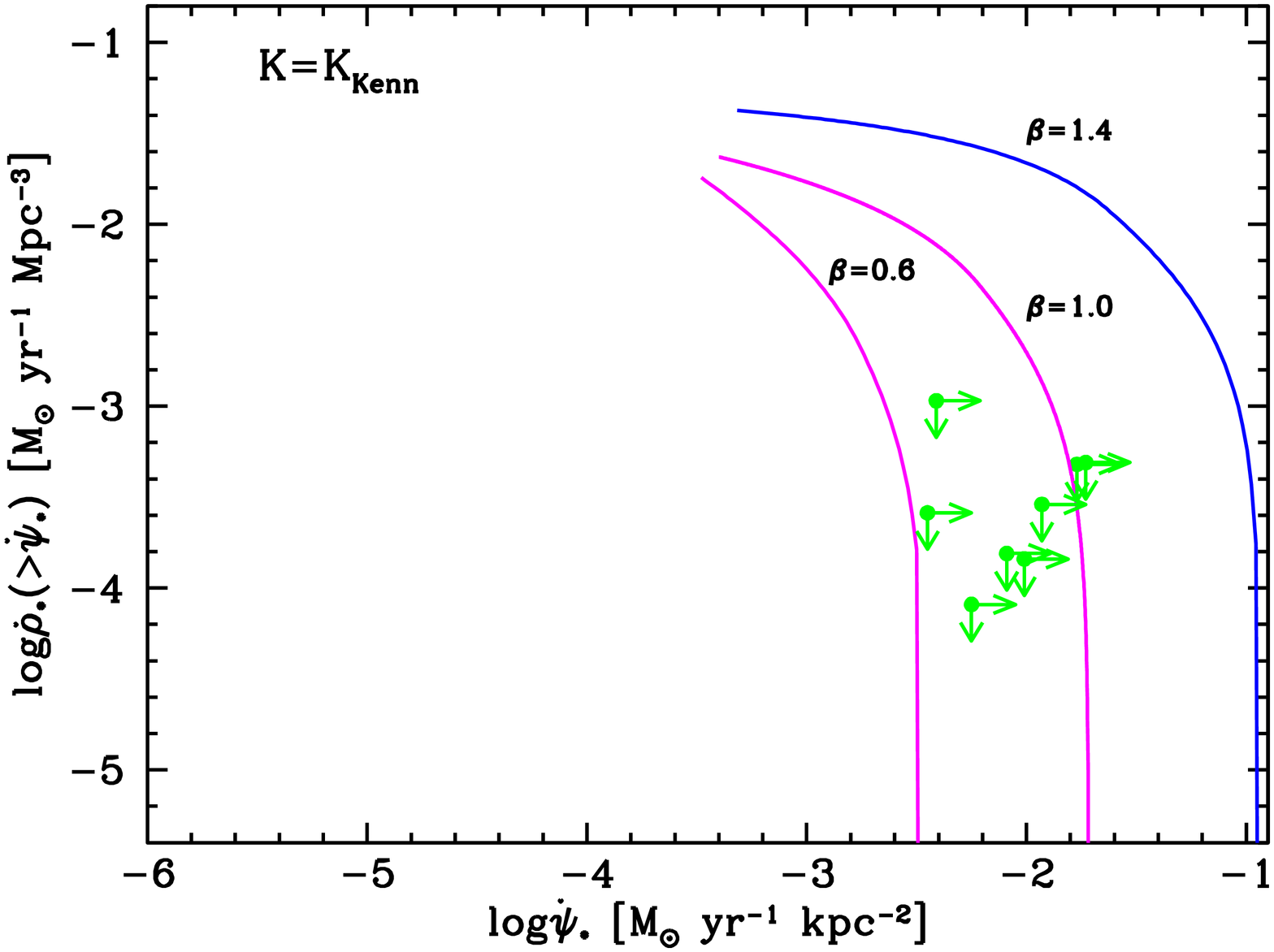}
\caption[]{Cumulative comoving SFR density versus 
{\ps}. Data points 
as  in Fig.~{\ref{fig:rhodotvN}}. Theoretical curves correspond
to $K$ = $K_{Kenn}$ for slopes {$\beta$}= 1.4, 1.0, and 0.6.}
\label{fig:rhodotvN_bet}
\end{figure}



The SFRs would be reduced if the critical surface density for the 
Kennicutt-Schmidt
law (see Eq.~{\ref{eq:Kenlaw}}) were higher in DLAs. The physical basis
for this follows from the Toomre instability criterion, which is
characterized by the parameter $Q$=
${\kappa}{\sigma_{g}}/{\pi}G{\mu}{m_{H}}N_{\perp}$ (Toomre 1964),  
where 
$\sigma_{g}$ is the gas velocity
dispersion, and $\kappa$ is the epicyclic frequency. 
The critical H I column density,
{\nh}$_{\perp}^{crit}$=${\kappa}{\sigma_{g}}/{\pi}G{\mu}{m_{H}}$, is
set by the condition $Q$=1:  
gas with $N_{\perp} >N^{crit}_{\perp}$ is unstable to
collapse and possible subsequent star formation, while gas with $N_{\perp}
{\le} N^{crit}_{\perp}$ is stable.
In a disk with a
flat rotation curve
$\kappa(r)={\sqrt 2}v_{rot}/r$, where $r$ is radial
distance from the center of the disk and $v_{rot}$ is the rotation speed.
Following Kauffmann (1996) and Mo {\etal} (1998) we assume the DLA
disk radius $r_{dla}{\approx} 0.1r_{200}$ where $r_{200}$ is
essentially the virial radius of the dark-matter halo enclosing the
disk and $r_{dla}$ is defined by the condition
$N(r_{dla})$=$N_{min}$=2{$\times$}10$^{20}$ cm$^{-2}$.  Adopting the spherical
collapse model for disk formation, one finds $v_{rot}/r_{200}=10H(z)$
(Mo {\etal} 1998) where $H(z)$ is the Hubble parameter. 
The presence of the $H(z)$ term
implies that {\nperp}($r$) at $z$ =3 must be at least a
factor of 5 higher than at $z$=0 for gas to form stars.
Specifically,
$N_{\perp}(r)$ must exceed $N_{\perp}^{crit}(r)$=
2.2{$\times$}10$^{21}$$(r_{dla}/r)$ cm$^{-2}$
for gas at radius  $r$  to be Toomre
unstable when  $z$ = 3. 
But  the double power-law form of {\fNX} rather indicates 
the gas is stable. 
For randomly oriented disks, the break of the
double power-law  occurs at $N_{d}$=$N_{0}$ where
$N_{0}$ is the maximum value of {\nperp}($r$):
in $\S$ 2.1 we found that $N_{d}$=3.16{$\times$}10$^{21}$ cm$^{-2}$.
Combining Eqs.~{\ref{eq:gdN}}
and 
~{\ref{eq:NumberDLA}}
 with the solution $g(N_{\perp},X)$
=$k_{3}(N_{\perp}/N_{0})^{-2}$, we find that  
$N_{\perp}(r)=N_{0}
{\Bigl [}1+20[(r^{2}-r_{0}^{2})/r_{dla}^{2}]{\Bigr ]}^{-1}$
for $r_{0} \ < r < \ r_{dla}$
where $r_{0}$ satisfies the condition $N_{\perp}(r_{0})=N_{0}$.
Comparison with 
$N_{\perp}^{crit}(r)$ shows that {\nperp}($r$) $<$ 
$N_{\perp}^{crit}(r)$ for all solutions in which {\nperp}($r_{dla}$)
$<$ $N_{min}$. 
Therefore, {\em in situ} star
formation  in DLAs may be suppressed because they are 
subcritical disks.


On the other hand the uncertainties in our model assumptions lead
to uncertainties in the values of $N_{\perp}^{crit}$. Moreover, uncertainties
in the slope
$\alpha_{4}$$\equiv$d$log${\fNX}/d$logN$ at $N$ $\ge$ $N_{d}$ indicate  that the
disk hypothesis, which predicts $\alpha_{4}$ = $-$3,
is not well established (see $\S$2.1). Therefore, gravitationally unstable
gas with $N$ $>$ 3.16{$\times$}10$^{21}$  cm$^{-2}$
may be present in DLAs. 
As a result, while cosmological
evolution may be an important factor in rendering
DLA gas subcritical, additional factors may be needed to
suppress star formation at  column densities up to 10$^{22}$ cm$^{-2}$.
One such factor may be the low molecular content
of DLAs: 
molecules are crucial for star formation since emission
from rotational transitions of CO and other molecules enables
clouds to cool below the atomic lower limit of $\sim$ 100 K
and to ultimately collapse and form stars. 
Whereas the median H$_{2}$ content in the diffuse neutral
gas of the Galaxy, $f_{H_{2}}$$\approx$10$^{-1}$ (Tumlinson
{\etal} 2002), Ledoux {\etal} (2003) find that $f_{H_{2}}$ $<$ 10$^{-6}$
in DLAs. Interestingly, the median values for
$f_{H_{2}}$ in the LMC and SMC are less
than 10$^{-4}$. Tumlinson {\etal} (2002) 
argue that the low molecular content of the
LMC and SMC  is likely due to low dust content
and relatively strong FUV radiation fields. Since
the low dust level suppresses H$_{2}$
formation on grains while the high FUV radiation level increases
the H$_{2}$ photodissociation rate,
higher values of $N$ are required for the gas to be
self-shielding against photodissociating radiation. Because DLAs have lower
dust-to-gas ratios than either the LMC or SMC but have similar FUV
radiation fields (WPG03),
the same
mechanisms may by responsible for the low molecular content
of DLAs.
Calculations of photodissociation equilibrium suggest that
$f_{H_{2}}$ would be suppressed at $N$ $<$ 10$^{22}$ cm$^{-2}$
in DLA gas (Tumlinson {\etal} 2002; Hirashita \& Ferrara 2005).
Krumholz \& McKee (2006) and Blitz \& Rosolowsky (2006) 
argue that {\ps} $\propto$ $f_{H_{2}}$, which would help to explain
the factor of ten or more
reduction in the efficiency of star formation in DLAs. 
Consequently, while gas in DLAs with  $N$ $>$ $N^{crit}$
$<$ 10$^{22}$ cm$^{-2}$ is Toomre unstable, it may be unable to
collapse to form stars. 
This is in contrast to the Galaxy 
and most metal-rich spirals studied by
Martin \& Kennicutt (20001) in which
the condition $N$ $>$ $N^{crit}$ normally results in star formation.
The explanation (R. C. Kennicutt 2006, priv. comm.) may be the result
of a coincidence between $N^{crit}$ and the column density at
which molecular gas becomes self shielding.
Because of its relatively high dust-to-gas ratio, this is lower
in the Galaxy (and most metal-rich spirals)
than in metal-poor objects such as the LMC, SMC (Tumlinson {\etal} 2002),
and in DLAs. This may
help to explain why the Toomre criterion alone appears to
be a sufficient condition for the onset of star formation in
most metal-rich spirals
(Martin \& Kennicutt 2001), but not in DLAs.

A further mechanism for cloud stabilization was suggested
by Ferguson {\etal} (1998) who argued that flaring of the H I disk in
the outer regions of spiral galaxies leads to suppression of star
formation: for a given gas surface density the increased
scale-height reduces the volume density, which helps to stabilize gas
against Jeans instability. 
Schaye (2004) argued that disks would be Toomre stable provided
the gas is a warm ($T$ $\sim$ 10$^{4}$ K) neutral medium and
unstable when the gas is a cold ($T$ $\sim$ 100K) neutral
medium.
But independent lines of evidence 
(e.g. WGP03; Howk {\etal} 2005) 
suggest that half of
the known sample of DLAs is comprised of cold ($T$ $\sim$ 100K) gas
with column densities exceeding the (5$-$20){$\times$}10$^{20}$ cm$^{-2}$
instability threshold. 
While this mechanism
needs to be investigated further, current evidence suggests
that phase transitions alone are not sufficient to trigger
star formation in  DLAs.
There may be other mechanisms that
do suppress star formation, but further discussion is beyond the scope of
this paper.

\subsection{Implications for Metal Production and Heat Input: the LBG
Connection}

The  UDF results place
a conservative upper limit on {\rhodot}
contributed by  DLAs with {\nh} $\ge$  $N_{min}$, which is
given by {\rhodot} $<$ 10$^{-2.4}$
{\smpympc} for $z$ $\approx$ 3 (see curve with
$K$=0.1{$\times$}$K_{Kenn}$ in Fig.~{\ref{fig:rhodotvN}}).
The new limit on {\rhodot} has several consequences. First, since the
metal production rate is proportional to {\rhodot}, one can compute
the DLA metalicity at a given redshift from the previous history of
{\rhodotz}. Nagamine {\etal} (2004b) used their numerical simulations,
which incorporated the Kennicutt-Schmidt law, to predict {\rhodot} =
10$^{-1}$ {\smpympc} and the metallicity [M/H]=$-$0.5 at $z$=3.  If we
scale their result to our upper limit on {\rhodot}, their chemical
evolution model predicts [M/H] $<$ $-$1.9.  By comparison the average
DLA metallicity at $z$ $\approx$ 3 is given by
[M/H]=$-$1.40{$\pm$}0.07 (Prochaska {\etal} 2003). Therefore, our
limit on {\rhodot} has shifted 
the problem of metal overproduction by a factor of ten in
DLAs (see WGP03) to one
of underproduction a factor of three.

Second, the new limit on {\rhodot} implies a gas heating rate
significantly lower than indicated by the measured cooling rate.  The
[C II] 158 {\micron} cooling rate per unit comoving volume inferred
from {\ciis} absorption in DLAs (Wolfe {\etal} 2004) is given by
\begin{equation}
{\cal C}={\Omega_{g}}{\rho_{crit}}{<{\ell_{c}}>}/{{\mu}m_{H}} 
\label{eq:coolcomov}
\cmma
\end{equation}

\noindent where
{$\Omega_{g}$} is the mass per unit comoving volume of neutral gas in
DLAs, 
{\lcav} is the average [C II] 158 {\micron} cooling rate per H atom,
and ${\mu}m_{H}$ is the average mass per particle.  Assuming
$\mu$=1.3, our data indicate
{$\cal C$}= (2$\pm$0.5){$\times$}10$^{38}$ ergs s$^{-1}$ Mpc$^{-3}$ in
the redshift interval $z$=[2.5,3.5]. Here we have ignored half the DLA
population with {\lc} $<$ 10$^{-27.1}$ ergs s$^{-1}$ H$^{-1}$ because
it is likely heated by background radiation alone (Wolfe 2005).  

To
balance cooling, Wolfe {\etal} (2003a [hereafter WPG03]) suggested the
grain photoelectric effect as a plausible heating mechanism. 
The comoving heating rate is given by

\begin{equation}
{\cal H}= \int {\phi}(L_{\nu})H(L_{\nu})dL_{\nu}
\label{eq:heatcomov1}
\cmma
\end{equation}

\noindent where ${\phi}(L_{\nu})$$dL_{\nu}$ is the comoving density
of DLAs with FUV luminosities in the interval ($L_{\nu},L_{\nu}$$+$d$L_{\nu}$).
The heating rate per DLA is given by

\begin{equation}
H(L_{\nu})= \int {\Gamma}({\bf r})n({\bf r})dV
\label{eq:heatpDLA}
\cmma
\end{equation}

\noindent where ${\Gamma}({\bf r})$ is the heating rate per H atom,
$n({\bf r})$ is the density of H atoms at displacement vector
${\bf r}$, and the integral
extends over the DLA volume. From the models of
Bakes \& Tielens (1994) and Weingartner \& Draine (2001),
WPG03 find
that $\Gamma$=10$^{-5}$${\kappa}{\epsilon}J_{\nu}({\bf r})$
where 
$\kappa$ is the dust-to-gas ratio, $\epsilon$ is the heating
efficiency, and $J_{\nu}({\bf r})$ is the FUV mean intensity.
Assuming DLAs are uniform disks with radius $R$ and scale-heights $h$,
and adopting the solution for $J_{\nu}$ (WPG03) one 
finds

\begin{equation}
{\cal H}={{{10^{-5}}
{{{\kappa}{\epsilon}{<N>} \over {8{\pi}}}[1+{\ln}(R/h)]}}}\int {\phi(L_{\nu})}L_{\nu}dL_{\nu} 
\label{eq:heatcomov2}
\cmma
\end{equation}


\noindent where 
 $<N>$ is the average H I column density of the DLA sample. 
Because {\rhodot}=1.25{$\times$} \\ 10$^{-28}$$\int {\phi}(L_{\nu})L_{\nu}dL_{\nu}$
(Madau {\etal} 1998),
the comoving heating rate {$\cal H$} is proportional to {\rhodot}.
Assuming parameters adopted by Wolfe {\etal} (2004), we find
that our upper limit on {\rhodot} implies {$\cal H$} $<$
4{$\times$}10$^{37}$ ergs s$^{-1}$ Mpc$^{-3}$.  Therefore, 
{\em in situ} star formation in DLAs is unlikely
to account for the energy input needed to balance cooling.

Consequently the new limits on {\rhodot} require an external source of
metals and heat input for DLAs with {\lc} $>$ 10$^{-27.1}$ ergs
s$^{-1}$ H$^{-1}$. 
This idea was first discussed by
WGP03 who suggested that centrally located
galaxy bulges could heat the absorbing gas. 
Wolfe (2005) then suggested LBGs as the most
natural bulge sources.
In this scenario the
FUV radiation originates in the LBGs, is attenuated by
their high dust content,  and then  
propagates unattenuated
throughout DLAs because of their low
dust content. The LBGs heat the surrounding DLA gas, 
with the same FUV radiation
observed by us to have {\rhodot}= 10$^{-1.7}$ {\smpympc} at $z$=3
(e.g.  Giavalisco {\etal}2004). From Eq.~{\ref{eq:heatcomov2}}, which
also applies for central point sources (WGP03), we find
the resulting LBG heating rate to be
{$\cal H$}=(3$\pm$2}){$\times$}10$^{38}$ erg s$^{-1}$ Mpc$^{-3}$,
where the large systematic errors stem from uncertainties in the dust
composition, photoelectric heating efficiency, geometry, etc.
Nevertheless this estimate suggests that embedded LBGs can explain the
heating rate inferred for DLAs with {\lc} $>$ 10$^{-27.1}$ ergs
s$^{-1}$ H$^{-1}$.

The physical association of high-$z$ DLAs with LBGs is supported by 
the following phenomena. First, Cooke {\etal} (2006) recently measured
the 3-D cross correlation function between DLAs and LBGs and found
good agreement between the DLA-LBG cross-correlation
amplitude and the LBG-LBG autocorrelation amplitude.
This implies a similarity in the dark-matter masses of the
two populations; i.e., a strong overlap between DLAs and LBGs. Further
evidence for this association comes from the
identification
of at least one high-$z$ DLA with an LBG (M$\o$ller {\etal} 2002). In this object
the FUV surface-brightness of the compact emitting regions is two
orders of magnitude higher than predicted by the Kennicutt-Schmidt law
applied to the H I column-density detected in absorption. This
suggests that star formation originates in compact, presumably dusty,
regions rarely detected in absorption, which are surrounded by the lower
column-density gas that gives rise to the DLA. This configuration
brings to mind the ``tadpole galaxies'' discussed in $\S$ 3.2 and indicates
that their star forming regions may also be embedded in DLA gas. The
low area covering factor and high column density implies that star
formation in DLAs occurs in dense molecular regions (see Zwaan \&
Prochaska 2006), which are replenished with gas from the surrounding
DLA.  By contrast, the DLA gas may be chemically enriched by P-cygni
winds emanating from the star-forming LBGs (Steidel {\etal}
2003). Other sources of metals may be pre-enriched gas or stars
accreted by the galaxy hosting the DLA. The latter processes may be
crucial for DLAs with {\lc} $<$ 10$^{-27.1}$ erg s$^{-1}$ H$^{-1}$,
which may not contain centrally located LBGs.

\section{SUMMARY and CONCLUDING REMARKS}

We searched for {\em in situ} star formation in the neutral gas
comprising DLAs by looking for low surface-brightness emission from
spatially extended objects in the Hubble Ultra Deep Field 
F606W image.  The search was designed to detect objects in the redshift
interval $z=[2.5,3.5]$ with linear {\em diameters} ranging between 
1.9 kpc and 31 kpc, which encompasses most model predictions published
so far.  The search 
was optimized using a smoothing kernel to match the predicted
size of the DLAs, ranging between an angular diameter of {\thetdla}=
0.25{\arcsec} and 4.0{\arcsec}.
The results are summarized as follows:

{$\bullet$} After eliminating compact objects with high
surface brightness, we found the number of extended sources to decrease
rapidly with increasing {\thetkern}: the implication is that
extended sources of low surface-brightness
emission rarely occur in the UDF (see Fig.~{\ref{fig:dlanumvthet}}).
Because photometric redshifts are unavailable,
these findings lead to upper limits on the  comoving density
of objects within the UDF with
$z$=[2.5,3.5].


 {$\bullet$} The upper limits on comoving density
lead to upper limits on comoving SFR density,
which are between factors of 30 to 100 lower
than predicted
by the Kennicutt-Schmidt law.  As a result, the projected SFRs per unit
area must be at least a factor 10 lower than the rates 
predicted by the distribution of neutral-gas
column densities in DLAs.
The upper limits are  quite general as they apply to objects with
angular diameters
encompassing the sizes of DLAs in {\lcdm} and all other
models suggested so far. 

{$\bullet$} We suggest that 
lower SFR efficiencies may partly result from an increase 
with redshift of the critical
surface density for Toomre instability. Such an increase
is due to cosmological evolution alone. This effect
causes the DLA gas to be Toomre stable, which
could remove the discrepancy between
the comoving SFR density predicted by the Kennicutt-Schmidt
law and our empirical upper limits. 
However, the uncertainties are such that
gravitational instabilities may be present in DLAs.
In that case we suggest that the
low molecular content of the gravitationally bound 
clouds with $N$ $<$ 10$^{22}$ cm$^{-2}$  prevents further
cooling and thus inhibits star formation.
The low molecular content of 
the gas may be related to the minimum column
density
required for molecule formation (Krumholz \& McKee 2005;
Blitz \& Rosolowsky 2006) since
molecule formation in objects with
low dust content and high FUV radiation
fields such as found in DLAs could require 
values of {\nh} exceeding 10$^{22}$ cm$^{-2}$ 
for shielding against photo-dissociating
radiation. The low area covering factor of such 
high column densities may be the reason for
the low molecular
content measured in DLAs. 

{$\bullet$} The upper limits on comoving SFR density  reduce
the rate of 
metal production predicted in DLAs. The resultant upper limits on
metallicity at $z$ $\sim$ 3 are lower
than the observed metallicities by a factor of three. While this removes
the previous problem of metal overproduction (by a
factor of ten)  in DLAs, it leads
to a new problem of metal underproduction.

{$\bullet$} The upper limits on comoving SFR density
result in a comoving grain photoelectric heating rate that is
significantly lower than the comoving cooling rate inferred 
from [C II] 158 {\micron} emission
rates for about half the DLA sample. Therefore, external sources of heating
are required.  We argue that the observed comoving SFRs of
LBGs with $R$ $<$ 27 and $z$=[2.5,3.5] indicate a heat input sufficient to 
balance cooling. 

{$\bullet$} We suggest a scenario in which
half the DLA population with
significant [C II] 158 {\micron} cooling is powered
by centrally located, compact LBGs.
We speculate that
DLA metal
enrichment may  be due to  P-cygni outflow from the
metal-enriched
LBGs. The half of the DLA population without
significant 
[C II] 158 {\micron} cooling is presumably heated
by background radiation alone. Without an LBG, the source of metals
for this DLA population is unclear.


Two  possible caveats come to mind.
First, since our limits do not apply to emission
from DLAs with {\thetdla} $<$ 0.25 {\arcsec}, it is
possible that for most DLAs at $z$ $\sim$ 3,  
$d_{dla}$ is  
less than 1.9 kpc 
and the SFRs follow the Kennicutt-Schmidt relation.
However, as discussed in $\S$ 3.1 all models
suggested so far predict that the bulk of DLAs will
have diameters larger than 1.9 kpc.

The second caveat stems from the possibility that
{\em in situ} star formation at the Kennicutt-Schmidt rate
occurs in DLA gas associated with high surface-brightness objects,
since such objects were automatically excluded from the survey.
We tested this hypothesis by searching for low surface-brightness
emission surrounding high surface-brightness objects. To enhance
the sensitivity of our search we smoothed the regions in the F606W
image surrounding selected targets using
kernels with 0.5 {\arcsec} $<$ {\thetkern} $<$ 2.0 {\arcsec}. 
In all cases the convolved images failed to show evidence for
outlying emission with fluxes comparable to the central object.
But it is difficult to draw conclusions from these data since the
redshifts of all but two galaxies, at $z$ = 2.593 (Szokoly et al.\ 2004)
and $z=3.797$ (Vanzella et al.\ 2006), respectively,
are unknown.
We found these two galaxies to consist of a compact core
(FWHM = 0.3
{\arcsec}) surrounded by diffusion emission, which exhibited maximum
contrast with respect to sky when  {\thetkern} = 0.5 {\arcsec}. 
In the case of the $z$=2.593 galaxy
the magnitude of the diffuse emission, $V$=30, while
the central core  has $V$ = 25; i.e.,
the regions surrounding this star-forming galaxy
contributes less than 1 $\%$ of the total SFR. 
Because {\rhodot} due to compact star forming galaxies is comparable to
{\rhodot}
predicted by the Kennicutt-Schmidt relation for DLAs, 
we would conclude that neutral gas
surrounding star-forming objects does not form stars according to the
Kennicutt-Schmidt relation if this galaxy is typical of its class.

Finally we compare our empirical limits with previous analyses that
apply the local Kennicutt--Schmidt law for evaluating {\rhodot} in DLAs.
First, Lanzetta {\etal} (2002) 
measured the frequency distribution of projected
SFR per unit area, $h$({\ps}), for galaxies
in the HDF in the redshift interval $z$=[0,10]. Their
measurements at $z$ $\sim$ 3 were sensitive to {\ps}
$\ge$ 10$^{-1.5}$ {\smpykpc}. To determine $h$({\ps}) at
lower values of {\ps}, they applied
the Kennicutt--Schmidt law to the 
H I column-density distribution function {\fNX} by using the
relation $h$({\ps})$d${\ps}=($H_{0}/c$){\fNX}$dN$. 
They then used Eq.~{\ref{eq:rhostardotN}} in the spherically
symmetric limit to determine {\rhodot}. 
These authors concluded that at $z\approx 3$ the bulk of SFR density 
resides in regions of 0.1 {\msolar} yr$^{-1}$ kpc$^{-2}$, where the majority
of LBG's are found.  They argued that DLAs of {\nh} $<5\times 10^{21}$ 
cm$^{-2}$ contribute an  insignificant fraction of the global
SFR density. This is inconsistent with the current form of
{\fNX} which when extrapolated
to
{\nh} $>>$ 2{$\times$}10$^{22}$ 
cm$^{-2}$, to include HSB objects such as LBGs,
predicts that 
more than 50 $\%$ of {\rhodot} is
contributed by 
DLAs with {\nh} $<5\times 10^{21}$ cm$^{-2}$,
even when $K$=$K_{Kenn}$/2.8
as suggested by Lanzetta {\etal} (2002).  
The discrepancy stems from
the different expressions used to compute {\fNX}: Lanzetta {\etal}
(2002) used  the then available expression for {\fNX} 
with  
slope, {$\alpha_3$}=$-$1.4,  that is flatter than
$\alpha_{3}$=$-$2.0 derived for the more
accurate SDSS function (PHW05) used here.
Because we have shown that the DLA  contribution
to {\rhodot} cannot be this large, 
the newer data do not support the 
suggestion of Lanzetta {\etal} (2002) that
SFRs predicted by application of the Kennicutt-Schmidt law
to DLAs
overlap  those derived directly from optical emission.
On the other hand, their analysis 
is consistent with our interpretation that LBGs provide
the bulk of the star formation at
$z$ $\sim$ 3.
Second, Hopkins {\etal} (2005) combined an updated version of {\fNX}
with the Kennicutt-Schmidt relation to predict
{\rhodot} in the spherically symmetric limit for 
the redshift interval $z$=[2.5,3.5]. 
These authors concluded that DLAs contribute roughly 50\% of the observed
SFR density at $z\sim 3$.
Again, the predicted {\rhodot} is significantly higher than our empirical
upper limits, and thus
their models are
ruled out by the UDF data.
The new upper limits on {\rhodot} are also
inconsistent with the predictions of the uniform-disk model
of WPG03 and WGP03. Their models
do not make use of the Kennicutt-
Schmidt law, but rather use the strength of 
{\ciis} absorption to 
infer the level of {\em in situ} star formation throughout
DLAs modeled as uniform disks. Because the predicted average 
projected SFR, $<${\ps}$>$=10$^{-2}$ {\smpykpc}, is above
our detection threshold, the factor of 100 disagreement between the predicted
{\rhodot} and our upper limits
rules out this model. 
On the other hand, the new upper limits
are consistent with the ``bulge model'' suggested by 
WGP03. In this model, star formation
does not occur in the gas detected in absorption, but rather
arises in a centrally located compact region, 
which heats
the surrounding disk gas by emitting FUV radiation.
The rough agreement between the comoving heating rate
supplied by LBGs and the comoving cooling rate of
DLAs suggests the bulges are in fact LBGs.

\acknowledgments

We wish to thank Eric Gawiser, Kim Griest, 
Andrei Kravtsov, and Jason X. Prochaska
for valuable discussions. AMW was partially supported by 
NSF grant AST 03-07824 and H-WC was partially supported by
NASA grant 
NNG06GC36G.

\end{document}